%% file: jhep2.tex
\documentclass[paper,nohyper]{JHEP3} % 10pt is ignored!
\usepackage[english]{babel}
\usepackage{cite}
\usepackage{wasysym}
\usepackage{amssymb}
\usepackage{latexsym}
\usepackage{graphicx}
\usepackage[dvips]{epsfig}

\input{defs}

\title{Theoretical uncertainties for measurements 
 of $\alpha_s$ from electroweak observables} 

\author{Hasko Stenzel \\
II.Physikalisches Institut\\
Justus-Liebig University of Giessen\\
Heinrich-Buff Ring 16, D-35392 Giessen, Germany\\
\email{Hasko.Stenzel@cern.ch}}

\abstract{
One of the most precise measurements of the strong coupling constant $\alpha_s(\mz)$ 
is obtained in the context of global analyses of precision electroweak data. 
This article reviews the sensitivity of different electroweak observables to $\alpha_s$ and 
describes the perturbative uncertainties related to missing higher orders.     
The complete renormalisation scale dependence for the relevant observables 
is calculated at next-to-next-to-leading order and a new method is presented 
to determine the corresponding perturbative uncertainty for measurements of 
$\alpha_s$ based on these observables.}
\preprint{}  % 

\begin{document}

\section{Introduction}
\label{sec:intro}
High precision measurements of the parameters of the Standard Model (SM) have been performed  
over the last 15 years in particular at LEP, SLC and TEVATRON. Cross sections, asymmetries,  
masses and widths of the electroweak (EW) gauge bosons have been determined with a relative accuracy 
of often better than one per mil. The measurements as a whole over-constrain the SM and allow 
for an internal consistency check. In a global fit to these data certain unknown parameters 
like the mass of Higgs boson can be determined, or other quantities not directly measured at LEP 
like the mass of the top quark can be inferred from the LEP data alone \cite{EWG}. The sensitivity 
of the electroweak data to these parameters arise from higher order corrections. 

The LEP and SLD experiments have carried out global SM fits to combined data from 
various experiments and determine five parameters simultaneously: the masses of the Z and Higgs 
bosons and of the top quark, the hadronic vacuum polarisation $\Delta \alpha^{(5)}_{had}$ and the 
strong coupling constant $\alpha_s(\mz)$ constitute one often used set of parameters. The strong 
coupling constant plays a special role in these fits. It is essentially determined by hadronic 
observables, for which complete next-to-next-to-leading oder (NNLO) 
calculations are available. The measurement of $\alpha_s$ benefits 
from third order $\cal O$$(\alpha_s^3)$ perturbative QCD calculation, 
which is not yet complete for other 
variables like jet rates, event shapes or fragmentation functions. The theoretical 
systematic uncertainties related to missing higher orders are expected to be smaller than 
for NLO-based determinations, even including all-orders resummation used 
for analyses of event-shape distributions \cite{Hasko}. 

A detailed analysis of the perturbative uncertainty associated to this kind of
measurement of $\alpha_s$ from electroweak observables is the purpose of this article. 
The renormalisation scale dependence of the theoretical predictions is taken as indicative 
for the systematic theoretical uncertainty originating from missing higher orders, keeping 
in mind that the real uncertainty may in fact be different and will be revealed only 
once the higher orders are actually calculated. Albeit this limitation, the renormalisation 
scale variation is commonly used in other processes and for other variables as well and 
may therefore at least be used for a relative comparison of different measurements.
 
Other methods have been proposed in the past to estimate the uncertainty for EW observables. 
Instead of a scale variation certain classes of higher oder terms 
where calculated \cite{Soper} in order to improve the convergence of the perturbation series 
by minimising renormalon effects, 
and the difference with respect to the standard NNLO result was taken as uncertainty. This 
procedure leads to a very small estimate of $\pm 0.0005$ for the uncertainty of $\alpha_s{\mz}$, 
which should be taken  
with care given that only a subset of the higher order terms have been calculated. 

The scale variation method was investigated in \cite{Bethke1}, where a parameterisation 
\cite{Bethke2}  
of $R_Z$, the ratio of hadronic to leptonic width of the $Z$ boson was used 
to derive an uncertainty of $+0.0028$ $-0.0005$ for $\alpha_s$. However, the 
parameterisation embodies only the the scale dependence of the massless NNLO 
part of the calculation but neglects the also scale dependent massive quark and 
mixed EW$\otimes$QCD corrections.       

The objective of this paper is the complete evaluation of the renormalisation 
scale dependence for all relevant observables and an analysis of the contribution 
from different classes of higher order correction to the systematic uncertainty    
defined by a scale variation.           

The experimental systematic uncertainties for $\alpha_s$ are studied in detail 
by the LEP experiments as well as the correlation between $\alpha_s$ and the other 
EW observables. Taking the leptonic cross section at the Z pole as observable from which 
$\alpha_s$ is determined, 
 a variation of the Higgs mass in the range from 
100 to 1000 GeV entails a change in $\alpha_s$ of about $0.0022$, which has to be 
compared to an experimental uncertainty of $\pm 0.0030$.      

This paper is organised as follows. In Section \ref{sec:ewobs} the EW observables 
are briefly presented, in Section \ref{sec:thpred} the theoretical predictions 
for the widths, the effective couplings and the final state QCD corrections incorporated in the 
radiator functions are discussed. In Section \ref{sec:uncert} 
the theoretical uncertainties of the observables arising from the renormalisation scale 
variation are evaluated. The scale dependence is used in Section \ref{sec:aserror} to assess 
the perturbative uncertainty for measurements of $\alpha_s$ extracted from global fits, 
supplemented by an experimental cross-check using test data for the EW observables.
The conclusion and summary are given in Section \ref{sec:conc}.   

\section{Electroweak observables}
\label{sec:ewobs}
The sensitivity of certain electroweak observables to $\alpha_s$ arises mainly through  
pure QCD corrections $\cal O$$(\alpha_s^3$) to the decay widths of the $Z$ boson into 
hadronic final states. 
In addition, mixed QCD$\otimes$EW corrections $\cal O$$(\alpha\alpha_s$) 
to the electroweak couplings give rise to a dependence of both hadronic and leptonic 
observables on $\alpha_s$. Numerically the former corrections amount to about three percent, 
while the mixed corrections, being suppressed by a factor of $\alpha$, are below one per mil. 
In practice only observables containing the hadronic or total widths significantly contribute 
to the determination of $\alpha_s$ 
\begin{equation}\label{defwid}
\Gamma_h=\sum_q\Gamma_q \; , \; \; \Gamma_q= \Gamma(Z\rightarrow q \bar q) \; , \; \; \Gamma_Z=\Gamma_h + \Gamma_l + \Gamma_\nu \; .
\end{equation}
Beyond the width itself the $R$ ratio 
\begin{equation}\label{defr}
R_Z=\frac{\Gamma_h}{\Gamma_l} \; , 
\end{equation} 
is of interest as the mixed corrections to the couplings cancel to a large extent 
in the ratio. 
For measurements of $\alpha_s$ the most important 
observables are the leptonic and hadronic pole cross sections
\begin{equation}\label{defsig}
\sigma_h^0 = 12 \pi \frac{\Gamma_e\Gamma_h}{M_Z^2\Gamma_Z^2} \; , 
\; \; \sigma_l^0 = 12 \pi \frac{\Gamma_e\Gamma_l}{M_Z^2\Gamma_Z^2} \; .
\end{equation}        
In particular $\sigma_l^0$ exhibits a good sensitivity to $\alpha_s$ which 
allows for a precise measurement of $\alpha_s$ from this observable alone \cite{EWG}. 
As the leptonic pole cross section $\sigma_l^0=\sigma_h^0/R_Z$ is not independent 
of the other variables, it is not included in the global analyses. 
Realistic observables are $\Gamma_Z$, $R_Z$ and $\sigma_h^0$  in the sense 
that they are independent observables with  
a substantial sensitivity to $\alpha_s$.  

The calculations presented in this paper are carried out throughout with the electroweak 
library ZFITTER version 6.36 \cite{zfitter}. If not stated otherwise, the following numerical input values are used:
\begin{eqnarray}\label{inputs}
\mzl & = &  91.1875 \; \mbox{GeV}\\ \nonumber 
m_t & = & 175 \; \mbox{GeV}\\ \nonumber
\mhl & = & 150 \; \mbox{GeV}\\ \nonumber
\Delta\alpha_{had}^{(5)}(\mzs) & = & 0.02761 \\ \nonumber
1/\alpha(0) & = & 137.0359895 \\ \nonumber
\alpha_s(\mz) & = & 0.1185
\end{eqnarray}

\section{Partial widths}\label{sec:thpred}
The dependence on $\alpha_s$ and on the renormalisation scale of the relevant EW observables 
through the widths are given in the following sections.
\subsection{Dependence of the widths on $\alpha_s$}\label{subsec:asdep}
The partial width of the Z decay to a pair of fermions can be cast into two different 
expressions for leptons and quarks in order to incorporate the different types of radiative 
corrections.  The width for lepton pairs $l=e, \mu, \tau$ is given by:  
\begin{eqnarray}\label{lwidth}
\Gamma_l & = & \Gamma_0 | \rZl|\sqrt{1-\frac{4m_l^2}{\mzs}}\left(1+\frac{3}{4}\frac{\alpha(\mzs)}{\pi}Q_l^2\right)\\ \nonumber
& & \times \left[\left(1+\frac{2m_l^2}{\mzs}\right)(1+|\gZl|^2)-\frac{6m_l^2}{\mzs}\right] \; ,
\end{eqnarray}
while for quark pairs $q=u, d, s, c, b$ another expression is used: 
\bq
\gqq=\Gamma_0\,N_C\,\big|\rZq\big|
\Bigl[
|\gZq|^2\,R^{\fq}_{\sss{V}}(\mzs) + R^{\fq}_{\sss{A}}(\mzs)
\Bigr] 
+\Delta_{_{\rm EW/QCD}}\;.
\label{defzwidthq}
\eq
%--
The basic width $\Gamma_0$ is the given by  
%--
\bq
\Gamma_0 = \frac{\gf\mzc}{24\srt\,\pi} = 82.945(7)\,\mbox{MeV}\, , \; \; N_C = 3 \; ,
\eq
and $N_C$ is a QCD colour factor.
In the case of leptons mass effect terms $\propto \frac{m_l^2}{\mzs}$ are explicitly 
taken into account in Eq.~\ref{lwidth}, for quarks these effects are embodied in the   
radiator functions $R_V^q$ and $R_A^q$, which 
also account for QCD and QED radiative corrections. The core of the sensitivity of the 
widths and related EW observables to $\alpha_s$ stems from the 
dependence of the radiator functions 
on $\alpha_s$, and their renormalisation scale dependence dominates the theoretical 
uncertainty for a measurement of $\alpha_s$. 
The dependence of the widths and derived observables is depicted in Fig.~\ref{fig:asdep}, 
where the change of theses quantities normalised to their value at $\alpha_s(\mz)=0.1185$ is 
shown. Since the leading term of the QCD correction is $1+\alpha_s/\pi$, the dependence 
of the widths on $\alpha_s$ is basically linear, as higher order terms are suppressed 
by powers of $\alpha_s/\pi$.
\setlength{\unitlength}{1mm}
\FIGURE[h!]
{\epsfig{file=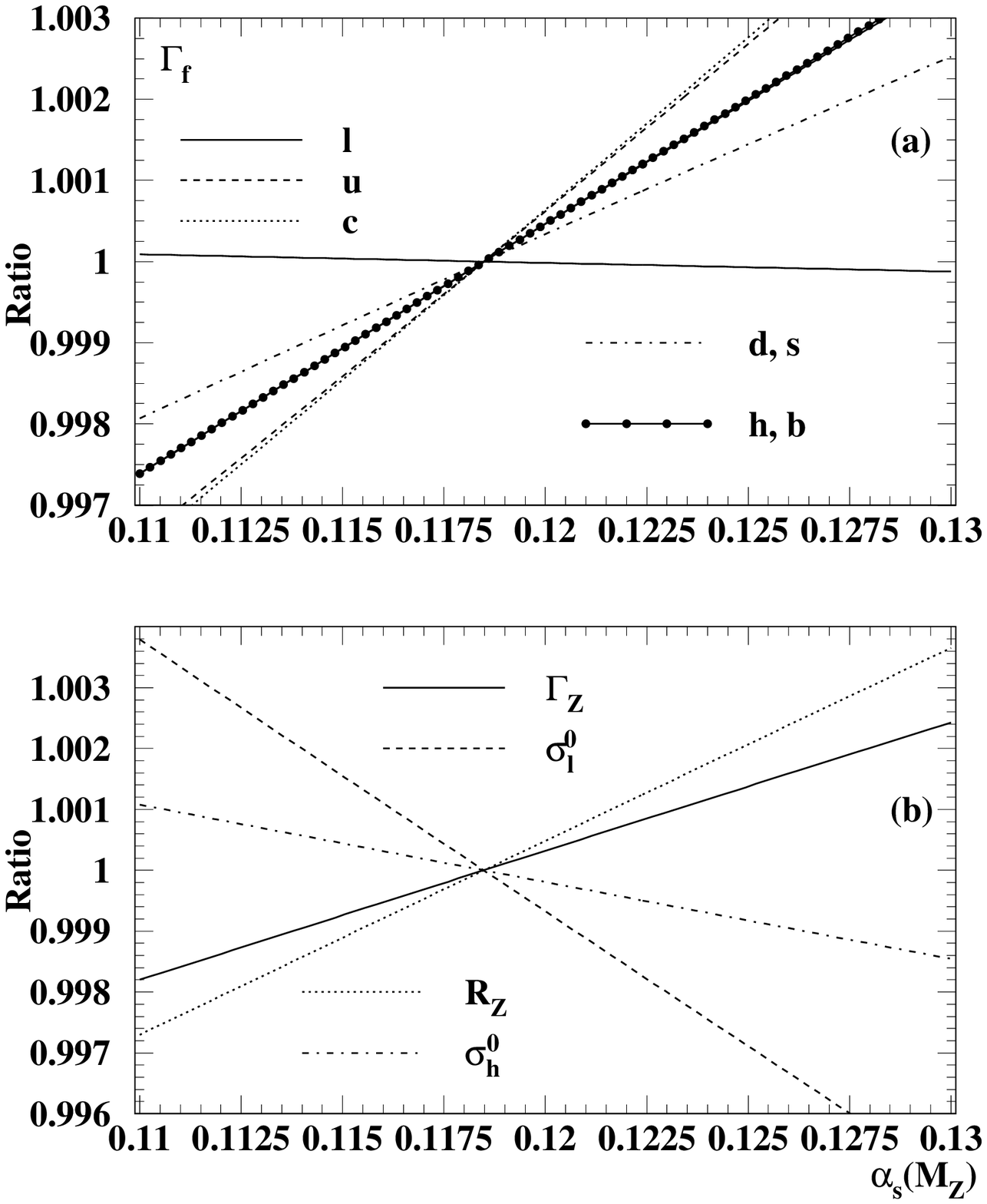,width=0.8\textwidth}
\caption{Dependence of the widths (a) and the EW observables (b) 
on $\alpha_s(\mz)$. Shown is the normalised ratio of the observable to its reference 
value at $\alpha_s(\mz)=0.1185$. The other SM parameters are kept fixed to their 
nominal values.\label{fig:asdep}}}

For the leptonic widths a very small change below 0.1 per mil is observed, induced 
by two-loop corrections to the effective EW couplings and to the vacuum polarisation contribution 
to $\alpha(s)$. The hadronic widths exhibit a stronger and opposite dependence resulting 
in a relative change between 4 and 6 per mil 
over the range of $\alpha_s$ between 0.11 and 0.13, this dependence is induced by 
the radiator functions. Among the hadronic widths there are clear differences between up- and 
down-type quarks, the d- and the s-quark dependences are identical. 
Furthermore, finite mass quark effects entail small differences between 
the u- and the c-quark, more visible between the d-quark and the b-quark. The b-quark 
behaviour is accidentally very close, but not identical to the sum over all flavours.

The dependence of the EW observables on $\alpha_s$ shown in Fig.~\ref{fig:asdep} is, as 
for the widths, almost linear and the relative change is between 2 and 8 per mil in 
the considered range. The observable 
with the strongest dependence is $\sigma_l^0$, where the radiator functions enter quadratically in 
the denominator.        

For quarks additional non-factorisable EW$\otimes$QCD corrections 
$\Delta_{_{\rm EW/QCD}}$ for the widths are not part of the radiator functions. These corrections 
are numerically very small (less than one per mil) and are taken as fixed numbers 
from \cite{czak1,czak2}
\begin{eqnarray}\label{czak}
\Delta_{_{\rm EW/QCD}} & = & -0.113 \; \mbox{MeV} \; \; \mbox{u-, c-quarks}\\ \nonumber
                         & = & -0.160 \; \mbox{MeV} \; \; \mbox{d-, s-quarks}\\ \nonumber
                         & = & -0.040 \; \mbox{MeV} \; \; \mbox{b-quarks} 
\end{eqnarray}
%--
The complex-valued variable $\rZf$ in Eqs.~\ref{lwidth}, \ref{defzwidthq} measures the overall strength of the neutral current 
interaction in the $f\bar f$ channel and the effective coupling $\gZf$ can be expressed 
in terms of the ratio of effective vector and axial-vector couplings    
%--
\bq
\Rvaz{\ff} = \frac{\vc{\ff}}{\ac{\ff}}
=1-4|\qf|\bigl(\kZdf{\ff}\siws+\Imsi{\ff}\bigr) \; ,
\label{varatiorez}
\eq
%--
where $\kappa_Z^f$ defines an effective mixing angle for flavour $f$ with $\siw = \sin^2\theta_W$ 
given by
\begin{equation}
\siws= 1-c_W^2\; , \; \; c_W^2 = \frac{M_W^2}{M_Z^2} \; .
\end{equation}
%---
The weak isospin $\tcif$ is $\pm 1/2$, the electric charge 
$\qf$ is $+\frac{2}{3} / -\frac{1}{3}$ for up-/down-type quarks and
the $\cal O$$(\alpha^2)$ term $\Imsi{\ff}$ originating from $\gamma\gamma$ and $Z\gamma$ 
polaristion operators is given by
\bq
\Imsi{\ff}=\alpha^2(\sman)\frac{35}{18}
\lrbr 1-\frac{8}{3}\Reb\bigl(\kZdf{\ff}\bigr)\siws\rrbr\; .
\label{addseffim}
\eq
The effective couplings of the Z decay $\kZdf{\ff}$ and $\rZf$  
incorporate radiative electroweak corrections up to two loops and their full expressions 
are given in \cite{zfitter}. The factorisable EW$\otimes$QCD corrections $\cal O$$(\alpha\alpha_s)$ 
shall be studied here, as they induce the $\alpha_s$ dependence to the effective couplings. 
 
%---
The running QED coupling denoted by $\alpha(\sman)$ is given by:
\begin{equation}\label{alphaem}
\alpha(s)=\frac{\alpha(0)}{1-\Delta\alpha_{had}^{(5)}(s)-\Delta\alpha_{lep}^{}(s)-\Delta\alpha^t(s)-
\Delta\alpha^{\alpha\alpha_s}(s)}\; .
\end{equation} 
The main contribution to the running coupling stems from the hadronic and leptonic 
vacuum polarisation. The leptonic part has been calculated at third order \cite{deltalep} 
\begin{equation} 
\Delta\alpha_{lep}^{}(\mzs)= 0.03149767 \; ,  
\end{equation} 
with negligible uncertainties. The hadronic contribution of the five light flavours 
is related via the dispersion relation to $R_{\gamma}$ (equivalent to $R_Z$ in the continuum) 
from which it can be extracted \cite{deltahad} 
\begin{equation} 
\Delta\alpha_{had}^{(5)}(\mzs)= 0.02761 \pm 0.0036 \; .  
\end{equation} 
The contribution from the top quark is small but depends on the 
top quark mass, for $m_t=175$ GeV 
\begin{equation} 
\Delta\alpha^{t}(\mzs)= -5.776 \; 10^{-5} \; .  
\end{equation} 
An explicit dependence on $\alpha_s$ appears in the 
$\cal O$$(\alpha\alpha_s)$ correction to $\alpha(s)$ representing gluonic 
insertions in $t\bar t$ loops \cite{Kniehl}. 
Of course the gluon exchange also 
occurs in light-quark loops and is accounted for in the experimental determination 
of $\Delta\alpha_{had}^{(5)}$. The correction for the top quark reads 
\begin{equation} 
\Delta\alpha^{t}(\mzs)= -\frac{\alpha\alpha_s}{\pi^2}\frac{4}{9} \left(\Reb\left(
\frac{V_1(r_Z)}{r_Z}\right)-4\zeta(3)+\frac{5}{6}\right) \; , \; \; 
r_Z=\frac{\mzl^2+i\epsilon}{4m_t^2} \; , \; \;  \zeta(3)= 1.2020569\; ,   
\end{equation} 
where the expression for $V_1(r)$ is given in \cite{Kniehl}. 
The numerical value is 
\begin{equation} 
\Delta\alpha^{\alpha\alpha_s}(s) = -1.02 \; 10^{-5} \; .
\end{equation}  
The dependence of $\alpha(s)$ on $\alpha_s$ is very weak, its relative 
change is about $10^{-6}$ for a variation of $\alpha_s(\mzl)$ between $0.11$ and $0.13$. 
%%%%%...
\subsection{Effective electroweak couplings}
The effective electroweak couplings $\rZf$ and $\kZf$ contain 
various self-energy terms, calculated for the EW part at NLO with two-loop 
corrections and at $\cal O$$(\alpha \alpha_s)$ for the mixed EW$\otimes$QCD corrections, 
including the terms leading in $m_t$ of the $\cal O$$(\alpha \alpha_s^2)$ contribution. 
The expressions for EW part are 
given in \cite{zfitter}, here only the terms relevant for the $\alpha_s$ dependence are 
summarised. In a convenient decomposition $\rZf$ and $\kZf$ are split into 
leading and remainder contributions, each being gauge invariant separately. The dominant 
leading term is re-summed to all orders in perturbation theory and the sub-leading 
remainder is calculated in fixed-order theory. 
The couplings $\rZf$ and $\kZf$ can be expanded in the following combination of leading 
and remainder terms: 
%---
\begin{eqnarray}\label{rhokappa}
\rZf & = &\frac{1+f_\alpha \left(\rho_{rem}^{f,G} + \rho_{rem}^{f,G\alpha_s}\right) +\rho_{rem}^{f,G^2}}
{1+\left(\hat\rho^G + \hat\rho^{G\alpha_s}\right)\left(1-\Delta r_{rem}^G -\Delta r_{rem}^{G\alpha_s}\right)} \;, \\ \nonumber
\kZf & = & \left[1+f_\alpha\left(\kappa_{rem}^{f,G}+\kappa_{rem}^{f,G\alpha_s}\right)+\kappa_{rem}^{f,G^2}\right]\\ 
 & & \times \, \left[1-\frac{c_W^2}{s_W^2}\left(\hat\rho^G+\hat\rho^{G\alpha_s}\right)\left(1-\Delta r_{rem}^G -\Delta r_{rem}^{G\alpha_s}\right)\right]
\; .
\end{eqnarray}
The transformation factor $f_\alpha$ accounts for the conversion of couplings $\alpha \rightarrow G_\mu$ \cite{zphys}
\begin{equation}
f_\alpha = \frac{\sqrt{2} G_\mu \mz s_W^2 c_W^2 }{\pi\alpha} \; .
\end{equation}  
The three-loop QCD correction to the $\rho$ parameter arising from top-quark loops is given by \cite{afmt} 
\begin{eqnarray}\label{leading}
\hat\rho^{G\alpha_s} & = & 3 x_t \left( c_{t1}\frac{\alpha_s(m_t)}{\pi} + c_{t2}\left(\frac{\alpha_s(m_t)}{\pi}\right)^2\right) \; , \\ \nonumber
& & x_t = \frac{G_\mu m_t^2}{8\pi^2\sqrt{2}} \; , \; \; c_{t1}=-2.86 \; , \; \; c_{t2}=-18.18 \; ,
\end{eqnarray}  
which also includes corrections to the term leading in $G_\mu m_t^2 \alpha_s^2$ derived in \cite{ChKS}. 
This is the dominant QCD correction to $\rZf$ and $\kZf$ and its renormalisation scale dependence  
determine the perturbative uncertainty for  $\rZf$ and $\kZf$ derived in Section \ref{sec:uncert}.

The expansions of $\rZf$ and $\kZf$ have also the remainders of the renormalisation parameter 
$\Delta r$ \cite{zfitter} in common, the component containing the QCD corrections is given by 
\begin{eqnarray}\label{delta_r}
\Delta r_{rem}^{G\alpha_s} & = & tb -tbl + 2cl \; ,\\
tb & = & \frac{\alpha\alpha_s(m_t)}{\pi^2} \, dr_{rem}\left(\mz, \mw, m_t^2\right) \; , \\
tbl & = & \frac{\alpha\alpha_s(m_t)}{4\pi^2}\frac{m_t^2}{\mw^2}\frac{\mz^2}{\mz^2-\mw^2}\left(\frac{1}{2} +\frac{\pi^2}{6}\right) \; ,  \\ \nonumber
cl & = & -\frac{\alpha\alpha_s(M_Z)}{4\pi^2}\frac{\mz^2\mw^2}{(\mz^2-\mw^2)^2}\log\left(\frac{\mw^2}{\mz^2}\right) \\ 
 & & \times \left[1+c_{l1}\frac{\alpha_s(M_Z)}{\pi}  + c_{l2} \left( \frac{\alpha_s(M_Z)}{\pi} \right)^2 \right] \; , \\\label{secondas2}   
 & & c_{l1}=1.409 \; , \; \; c_{l2}=-12.805 \; .
\end{eqnarray}
The QCD correction for the flavour-dependent remainder of $\rZf$ is
\begin{eqnarray}\label{rho_rem}
 \rho_{rem}^{f,G\alpha_s} & = & \rho^{QCD} + tbl \; , \\ \nonumber
 \rho^{QCD} & = & \frac{\alpha\alpha_s(m_t)}{\pi^2}\, d\rho_{rem}\left(\mz, \mw, m_t^2\right) \\ 
            &   & + \frac{\alpha\alpha_s(M_Z)}{\pi^2} \, \frac{V_T^2(t)+V_T^2(b)+2}{8s_W^2 c_W^2} \; , \\
  V_T(q) & = & 1-4Q_q s_W^2
\end{eqnarray}
The expression for $\kZf$ contains another QCD correction for the remainder 
\begin{eqnarray}\label{kappa_rem}
\kappa_{rem}^{f,G\alpha_s} & = & \kappa^{QCD} + \frac{M_W^2}{M_Z^2 -M_W^2}tbl - 3 x_t c_{a2}
\left(\frac{\alpha_s(m_t)}{\pi}\right)^2 \; , \\
 & &  c_{a2} = 0.644 \; , \\
\kappa^{QCD} & = &  \frac{\alpha\alpha_s(m_t)}{\pi^2} \, d\kappa_{rem}\left(\mz, \mw, m_t^2\right) \\
  & & +  \frac{\alpha\alpha_s(M_Z)}{\pi^2} \, \frac{c_W^2}{2s_W^4} \log c_W^2
\end{eqnarray}
The remainder functions $dr_{rem}, d\rho_{rem}$ and $d\kappa_{rem}$ describing the $\cal O$$(\alpha\alpha_s) $ contribution to the bosonic self-energies have been derived analytically in \cite{Kniehl}.
\setlength{\unitlength}{1mm}
\FIGURE[h!]{\epsfig{file=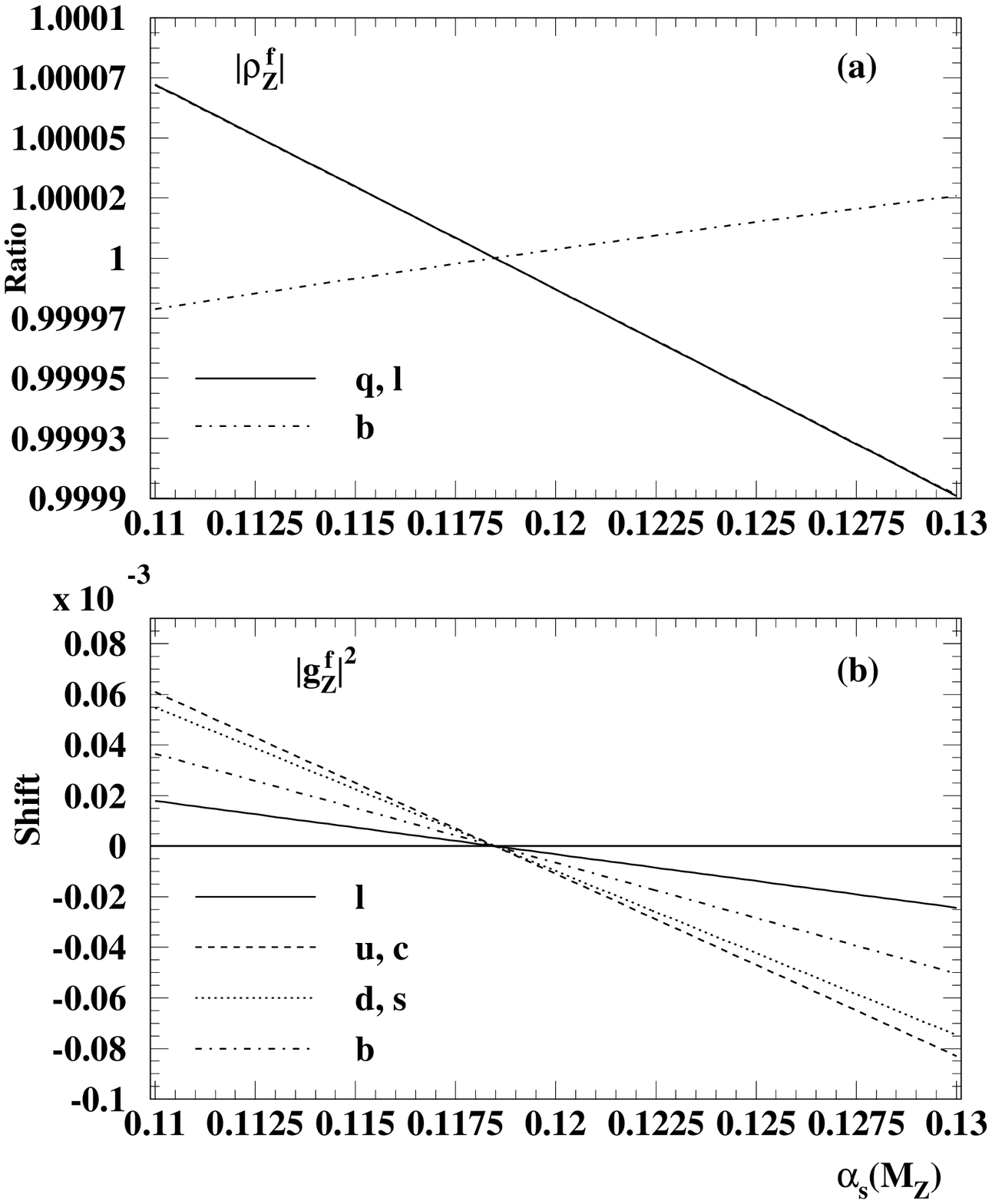, width=0.8\textwidth}
\caption{\small Dependence of the $|\rZf|$ (a) and $|\gZf|^2$ (b) 
on $\alpha_s(\mz)$. For $|\rZf|$ the normalised ratio and for $|\gZf|^2$ the shift of the coupling to its reference 
value at $\alpha_s(\mz)=0.1185$ is shown.\label{fig:as_roge}}}

In the case of b-quarks two additional one-loop vertex diagrams, absent for light quarks, are 
generated by the large mass splitting 
between the t- and the b-quark and contribute to the widths $\Gamma_b$ \cite{zbb}. These corrections 
entail a modification of the $Zb\bar b$ decay amplitude form factor \cite{zfitter}, which in turn 
affects the effective couplings. If $\rho^\prime$ and $\kappa^\prime$ denote the modified couplings 
given in \cite{zfitter},  
then the corrected couplings for the b quark are obtained by
\begin{eqnarray}\label{eq:taub} 
\rho_Z^b & = & \rho^\prime\left(1+\tau_b\right)^2 \; , \\ 
\kappa_Z^b & = & \frac{\kappa^\prime}{1+\tau_b} \; ,  \\
\tau_b & = & -2 x_t\left( 1 - \frac{\pi}{3}\alpha_s(m_t)+x_t\tau_2 \, \frac{m_t^2}{M_H^2}\right) \; , 
\end{eqnarray}
where the term in $\alpha_s(m_t)$ was obtained by \cite{ftjr} and the expression for 
$\tau_2$ can be found in \cite{taub2}.  

The dependence of the couplings on $\alpha_s$ is shown in Fig.\ref{fig:as_roge} for the absolute 
value of $\rZf$ and the squared module of $\gZf$, these are the relevant terms for 
the widths according to Eqs.~\ref{lwidth}, \ref{defzwidthq}.   
The dependence of $\rZf$ on $\alpha_s$ is weak, the relative change in the 
considered range is 
below $2 \cdot 10^{-4}$. All quark flavours and the leptons exhibit practically the same 
$\alpha_s$ dependence, except the b-quark for which it is even weaker and opposite. 

The representation of 
the relative change as ratio is inadequate for $|\gZf|^2$ since its absolute value is rather 
small, e.g. for leptons $|\gZl|^2 = 0.00555$ and $|\rZl|=1.00517$ at $\alpha_s(\mz)=0.1185$. 
Instead the shift defined by $|\gZf|^2(\alpha_s) -|\gZf|^2(\alpha_s=0.1185)$ is a better 
indicator for the dependence on $\alpha_s$. The absolute change of $|\gZf|^2$ is between 
$1.4 \cdot 10^{-4}$ for up-type quarks and $0.4 \cdot 10^{-4}$ for leptons. 
This translates into a relative change much larger for leptons than for quarks, 
given the small size 
of $|\gZl|^2$ for the leptons.  

The observed $\alpha_s$ dependence of the leptonic widths shown in Fig.~\ref{fig:asdep}
is dominated by dependence of $\rZf$ on $\alpha_s$. In the case of quarks, however, the 
couplings contribution to the $\alpha_s$ dependence of the widths is sub-leading, the 
main properties are determined by the QCD final state corrections in the radiator functions.  

%%%

\subsection{Radiator functions}\label{radiator}
Final state QCD and QED vector and axial vector corrections to the quarkonic widths 
Eq.~\ref{defzwidthq} are embodied in the radiator functions 
\begin{eqnarray}
R^{\fq}_{\sss{V}}(\sman)&=& 1 + \frac{3}{4} Q^2_q \frac{\alpha(\sman)}{\pi}
              +\frac{\als(\sman)}{\pi}
              -\frac{1}{4}Q^2_q\frac{\alpha(\sman)}{\pi}\frac{\als(\sman)}{\pi}
%             +{\cal O}(\alpha^2) 
\nll
        & &   + \left[C_{02}+C^t_2\left(\frac{s}{\mts}\right)\right]
                \left(\frac{\als(\sman)}{\pi}\right)^2         
              + C_{03}\left(\frac{\als(\sman)}{\pi}\right)^3
%             + {\cal O}(\alpha^4_{_S})  
\nll
        & &   + \frac{\mcS(\sman)+\mbS(\sman)}{s} C_{23}
                                \left(\frac{\als(\sman)}{\pi}\right)^3
\nll
        & &   + \frac{\mqS(\sman)}{s} \Biggl[ 
                        C^V_{21}      \frac{\als(\sman)}{\pi}
                      + C^V_{22}\left(\frac{\als(\sman)}{\pi}\right)^2
                      + C^V_{23}\left(\frac{\als(\sman)}{\pi}\right)^3
                                  \Biggr]
\nll
   & &+\frac{\mcQ(\sman)}{\smans}\left[ C_{42}-\ln\frac{\mcS(\sman)}{s}\right]
                                \left(\frac{\als(\sman)}{\pi}\right)^2 
      +\frac{\mbQ(\sman)}{\smans}\left[ C_{42}-\ln\frac{\mbS(\sman)}{s}\right]
                                \left(\frac{\als(\sman)}{\pi}\right)^2 
\nll 
        & &   + \frac{\mqQ(\sman)}{\smans} \Biggl\{
                       C^V_{41}       \frac{\als(\sman)}{\pi}
                 + \left[C^V_{42}+C^{V,L}_{42}\ln\frac{\mqS(\sman)}{s}\right]
                                \left(\frac{\als(\sman)}{\pi}\right)^2
                                    \Biggr\}
\nll
        & &   +12\frac{\mqpQ(\sman)}{\smans}
                                \left(\frac{\als(\sman)}{\pi}\right)^2
              -\frac{\mqX(\sman)}{s^3}\Biggl\{8+\frac{16}{27}
               \left[155+6\ln\frac{\mqS(\sman)}{s}\right]
                                      \frac{\als(\sman)}{\pi}\Biggr\},
\label{rvfact}
\end{eqnarray}
\begin{eqnarray}
R^{\fq}_{\sss{A}}(\sman)&=&1 + \frac{3}{4} Q^2_q \frac{\alpha(\sman)}{\pi}
             + \frac{\als(\sman)}{\pi}
             - \frac{1}{4}Q^2_q\frac{\alpha(\sman)}{\pi}\frac{\als(\sman)}{\pi}
%            + {\cal O}(\alpha^2) 
\nll
        & & + \left[C_{02}+C^t_2\left(\frac{s}{\mts}\right)
            - \left(2I^{(3)}_q\right)
              {\cal I}^{(2)}\left(\frac{s}{\mts}\right)\right]
              \left(\frac{\als(\sman)}{\pi}\right)^2         
\nll
        & & + \left[C_{03} - 
              \left(2I^{(3)}_q\right) 
              {\cal I}^{(3)}\left(\frac{s}{\mts}\right)\right]
              \left(\frac{\als(\sman)}{\pi}\right)^3
%           + {\cal O}(\alpha^4_s)
\nll
        & & + \frac{\mcS(\sman)+\mbS(\sman)}{s} C_{23}
                                \left(\frac{\als(\sman)}{\pi}\right)^3
            + \frac{\mqS(\sman)}{s} \Biggl[C^A_{20} 
                      + C^A_{21}      \frac{\als(\sman)}{\pi}
                      + C^A_{22}\left(\frac{\als(\sman)}{\pi}\right)^2
\nll
        & & + 6\left(3 + \ln\frac{\mts}{s}\right)
                                \left(\frac{\als(\sman)}{\pi}\right)^2 
                      + C^A_{23}\left(\frac{\als(\sman)}{\pi}\right)^3
                                \Biggr]
\nll
        & & - 10\frac{\mqS(\sman)}{\mts}
              \left[\frac{8}{81}+\frac{1}{54}\ln\frac{\mts}{s}\right]
                                \left(\frac{\als(\sman)}{\pi}\right)^2
\nll
& & + \frac{\mcQ(\sman)}{\smans}\left[ C_{42}-\ln\frac{\mcS(\sman)}{s} \right]
                                \left(\frac{\als(\sman)}{\pi}\right)^2
    + \frac{\mbQ(\sman)}{\smans}\left[ C_{42}-\ln\frac{\mbS(\sman)}{s} \right]
                                \left(\frac{\als(\sman)}{\pi}\right)^2
\nll 
        & &   + \frac{\mqQ(\sman)}{\smans} \Biggl\{C^A_{40} 
              + C^A_{41}       \frac{\als(\sman)}{\pi}
              + \left[C^A_{42}+C^{A,L}_{42}\ln\frac{\mqS(\sman)}{s}\right]
                                \left(\frac{\als(\sman)}{\pi}\right)^2
                                    \Biggr\}
\nll
        & &   -12\frac{\mqpQ(\sman)}{\smans}
                                \left(\frac{\als(\sman)}{\pi}\right)^2.    
\label{rafact}
\end{eqnarray}
Finite mass corrections are retained only for the b- and c-quark, i.e. $m_q=0$ for q$=$u,d,s, 
and the terms $m_q(s)$ represent the running quark masses in the $\rm\overline{MS}$ scheme. 
The term $m^\prime_q$ denotes the other quark mass in doublet, it is $m_c$ for q$=$b and $m_b$ for 
q$=$c. The different terms of Eq.~\ref{rvfact} and Eq.~\ref{rafact} and their coefficients can be 
organised in the following classes of corrections.\\ 
\noindent
Massless non-singlet corrections
\cite{Chetyrkin:1979bj,Dine:1979qh,Celmaster:1980xr,Gorishnii:1991hw}:
\bqa
C_{02} &=&  \frac{365}{24}-11\,\ztri
                         +\left[-\frac{11}{12}+\frac{2}{3}\,\ztri\right]\nf,
\\ \nll
C_{03} &=&  \frac{87029}{288}-\frac{121}{8}\,\ztwo
                         -\frac{1103}{4}\,\ztri +\frac{275}{6}\,\zfiv
\nll     
       & & +\left[-\frac{7847}{216}+\frac{11}{6}\,\ztwo
          +\frac{262}{9}\,\ztri-\frac{25}{9}\,\zfiv\right]\nf
\nll
       & & +\left[\frac{151}{162}-\frac{1}{18}\,\ztwo
          -\frac{19}{27}\,\ztri\right]\nfS,
\label{mass0}
\eqa
%---
with the number of active flavours $\nf$.\\
\noindent 
Quadratic massive corrections \cite{Chetyrkin:1994js3}:
\bqa
C_{23}&=&
-80+60\,\ztri+\left[\frac{32}{9}-\frac{8}{3}\,\ztri\right]\nf,
\\ \nll
C^V_{21}&=&  12,
\\ \nll
C^V_{22}&=& \frac{253}{2} - \frac{13}{3}\nf,
\\ \nll
C^V_{23}&=&  
2522-\frac{855}{2}\,\ztwo+\frac{310}{3}\,\ztri-\frac{5225}{6}\,\zfiv
\nll 
        & &+\left[-\frac{4942}{27}+34\,\ztwo
           -\frac{394}{27}\,\ztri+\frac{1045}{27}\,\zfiv\right]\nf
            +\left[\frac{125}{54}-\frac{2}{3}\,\ztwo\right]\nfS,\qquad
\\ \nll
C^A_{20}&=& -6,
\\ \nll
C^A_{21}&=& -22,
\\ \nll
C^A_{22}&=& -\frac{8221}{24}+57\,\ztwo+117\,\ztri
           +\left[\frac{151}{12}-2\,\ztwo-4\,\ztri\right]\nf,
\\ \nll
C^A_{23}&=& -\frac{4544045}{864}+1340\,\ztwo+\frac{118915}{36}\,\ztri
                                -127\,\zfiv
\nll 
        & &+\left[\frac{71621}{162}-\frac{209}{2}\,\ztwo-216\,\ztri
                                +5\zfor+55\,\zfiv\right]\nf
\nll 
        & &+\left[-\frac{13171}{1944}+\frac{16}{9}\,\ztwo
                                +\frac{26}{9}\,\ztri\right]\nfS;
\label{mass2}
\eqa
%---
\noindent 
Quartic massive corrections:
%---
\bqa
C_{42}     &=& \frac{13}{3}-4\,\ztri,
\\ \nll
C^V_{40}&=&-6,
\\ \nll 
C^V_{41}&=&-22,
\\ \nll
C^V_{42}&=&-\frac{3029}{12}+162\,\ztwo+112\,\ztri
           +\left[\frac{143}{18}-4\,\ztwo-\frac{8}{3}\,\ztri\right]\nf,
\\ \nll
C^{V,L}_{42}&=&-\frac{11}{2}+\frac{1}{3}\nf,
\\ \nll
C^A_{40}&=&6,
\\ \nll
C^A_{41}&=&10,
\\ \nll
C^A_{42}&=& \frac{3389}{12}-162\,\ztwo-220\,\ztri
           +\left[-\frac{41}{6}+4\,\ztwo+\frac{16}{3}\,\ztri\right]\nf,
\\ \nll
C^{A,L}_{42} &=& \frac{77}{2}-\frac{7}{3}\nf;
\label{mass4}
\eqa
%---
\noindent
Power suppressed t-quark mass correction:
%---
\bqa
C^t_2(x)&=&x\left(\frac{44}{675} - \frac{2}{135}\ln x \right);
\label{powsup}
\eqa
\noindent 
Singlet axial corrections:
\bqa
{\cal I}^{(2)}(x) &=&  - \frac{37}{12} + \ln x + \frac{7}{81}x
                       + \mbox{${\tt 0.0132}$}x^2,
\\ \nll   
{\cal I}^{(3)}(x) &=& - \frac{5075}{216} 
                      + \frac {23}{6}\,\ztwo + \,\ztri + \frac{67}{18}\ln x
                      + \frac{23}{12} \ln^2 x.
\eqa
%---
Here, the Riemann Zeta function  $\zeta$ is defined by
\begin{equation}
\zeta(x) = \sum_{n=1}^\infty n^{-x}
\end{equation}
with particular values 
\begin{equation}
\zeta(2)=1.6449341 \; , \; \;  \zeta(3)=1.2020569 \; , \; \;  \zeta(5)=1.0369278\; . 
\end{equation}
The evolution of the radiator functions with $\alpha_s$ is shown in Fig.~\ref{fig:as_rarv}.~
\setlength{\unitlength}{1mm}
\FIGURE[h!]
{\epsfig{file=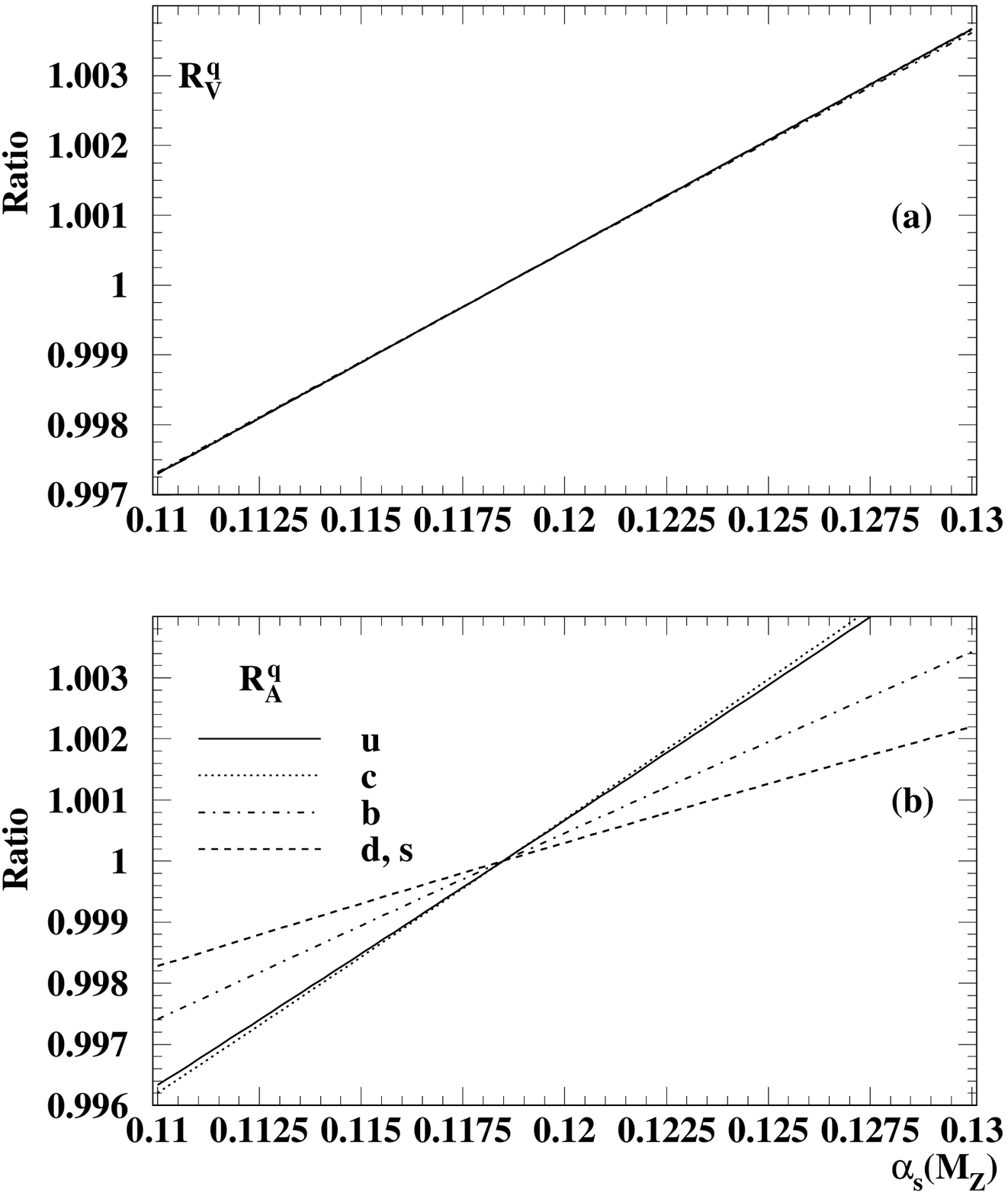 , width=0.8\textwidth}
\caption{\small Dependence of the radiator functions $R^{\fq}_{\sss{V}}$ (a) and 
$R^{\fq}_{\sss{A}}$ (b) 
on $\alpha_s(\mz)$. The ratio of the radiator function at a given value of $\alpha_s(\mz)$ 
and the reference value at $\alpha_s(\mz)=0.1185$ is shown.\label{fig:as_rarv}}}
The vector radiator functions for the different flavours are very similar and can barely be 
distinguished. Their dependence on $\alpha_s$ is almost linear and they are increasing by $0.6 \%$ for 
a change of $\alpha_s$ from $0.11$ to $0.13$. The axial-vector radiator function 
exhibits a prominent flavour dependence. For up-type quarks $R^{\fq}_{\sss{A}}$ increases by 
$0.8 \%$, 
for down-type quarks the change of the axial-vector radiator function is only of $0.4 \%$. 
The third component of the weak isospin $I^{(3)}_q$, present in the singlet axial component 
of $R^{\fq}_{\sss{A}}$, generates the up/down-quark difference. On top of this difference  
the b-quark mass corrections are important and entail a change of 
$R^{b}_{\sss{A}}$ of $0.6 \%$.     

%%%%%%%%%%%

\subsubsection{Running quark masses}
The running quark masses $m_q(s)$ in the $\overline{\rm MS}$ scheme are related to the 
fixed-valued pole masses $M_q$. For the c-quark at $s=M_s^2$ and $n_f=4$ it follows 
\bqa
\Mc &=& \mc(\McS) \Biggl\{1
   +\left[\frac{4}{3}+\ln\frac{\McS}{\mcS(\McS)}\right]\frac{\als(\McS)}{\pi}
\nll
&&+~\Biggl[K_c+\left(\frac{173}{24}-\frac{13}{36}\nf\right)
\ln\frac{\McS}{\mcS(\McS)}
+\left(\frac{15} {8} -\frac{1} {12}\nf\right)\ln^2\frac{\McS}{\mcS(\McS)}
\nll
&& +~\frac{4}{3} \Xi\left(\frac{\ms(\McS)}{\mc(\McS)}\right)\Biggr]
\left(\frac{\als(\McS)}{\pi}\right)^2\Biggr\},
\label{mcrun1}
\eqa
with 
\bqa
K_c&=&  \frac{2905}{288}+\frac{1}{3}\Biggl[7+2\ln(2)\Biggr]\,\ztwo
     -\frac{1}{6}\,\ztri
     -\frac{1}{3}\Biggl[\frac{71}{48}+\,\ztwo\Biggr]\nf,
\\
\Xi\left(x\right)
& = &
 \frac{\pi^2}{8}x-0.597 x^2 + 0.23 x^3.
\eqa
The running mass is evolved according to the renormalisation group equation 
from the scale of the pole mass $M_c^2$ to the scale of the process $s$ in a two-step 
evolution $M_c^2 \rightarrow M_b^2 \rightarrow s$:
%---
\bqa
\label{runmasc}
\mc(\sman) & = & \mc(\McS)
\Biggl[\frac{\als(\MbS)}{\als(\McS)}\Biggr]^{\gamma_{0}^{(4)}/\beta_{0}^{(4)}}
\Biggl\{1 + C_{1}(4)
\Biggl[\frac{\als(\MbS)}{\pi} - \frac{\als(\McS)}{\pi}\Biggr]
\\
& &  + \frac{1}{2} C^2_{1}(4)
\Biggl[\frac{\als(\MbS)}{\pi} - \frac{\als(\McS)}{\pi}\Biggr]^2
    + \frac{1}{2} C_{2}(4)
\Biggl[\Biggl(\frac{\als(\MbS)}{\pi}\Biggr)^2
     - \Biggl(\frac{\als(\McS)}{\pi}\Biggr)^2\Biggr]
\Biggr\}\qquad\qquad
\nll 
& &\times
\Biggl[\frac{\als(\sman)}{\als(\MbS)}\Biggr]^{\gamma_0^{(5)}/\beta_0^{(5)}}
\Biggl\{1 + C_1(5) 
\Biggl[\frac{\als(\sman)}{\pi} - \frac{\als(\MbS)}{\pi} \Biggr]
\nll
& &  + \frac{1}{2} C^2_{1}(5) 
\Biggl[\frac{\als(\sman)}{\pi} - \frac{\als(\MbS)}{\pi} \Biggr]^2
     + \frac{1}{2} C_2(5) 
\Biggl[\Biggl(\frac{\als(\sman)}  {\pi}\Biggr)^2
     - \Biggl(\frac{\als(\MbS)}{\pi}\Biggr)^2 \Biggr]
\Biggr\}.
\nonumber
\eqa
For the running b-quark mass the same procedure is applied with a single 
evolution from $M_b^2$ to $s$. The coefficients in Eq.~\ref{runmasc} are given by:
%---
\bqa
C_{1}(\nf) & = & 
\displaystyle{ \frac{\gamma_{1}^{(\nf)}}{\beta_{0}^{(\nf)}} 
             - \frac{\beta_{1} ^{(\nf)}
               \gamma_{0}^{(\nf)}}{\left(\beta_{0}^{(\nf)}\right)^2}}\,,
\\
\nl
C_{2}(\nf) & = & 
\displaystyle{ \frac{\gamma_{2}^{(\nf)}}{\beta_{0}^{(\nf)}} 
             - \frac{\beta_{1} ^{(\nf)}
               \gamma_{1}^{(\nf)}}{\left(\beta_{0}^{(\nf)}\right)^2}  
             - \frac{\beta_{2} ^{(\nf)}
               \gamma_{0}^{(\nf)}}{\left(\beta_{0}^{(\nf)}\right)^2}  
             +
             \frac{\left(\beta_{1}^{(\nf)}\right)^2\gamma_{0}^{(\nf)}}
{\left(\beta_{0}^{(\nf)}\right)^3}}\,.  
\eqa
%---
The coefficients of the Beta and Gamma functions are:
%---
\bqa
\beta_{0}^{(\nf)} & = & 
\displaystyle{ \frac{1}{4}  \left(11  -\frac{2}{3}   \nf \right) },
\\
\nl
\beta_{1}^{(\nf)} & = & 
\displaystyle{ \frac{1}{16} \left(102 -\frac{38}{3}  \nf \right) },
\\
\nl
\beta_{2}^{(\nf)} & = & 
\displaystyle{ \frac{1}{64} \left(\frac{2857}{2}-\frac{5033}{18}\nf 
              +\frac{325}{54} \nfS \right) },
\\
\nl
\gamma_{0}^{(\nf)} & = &  1,
\\
\nl
\gamma_{1}^{(\nf)} & = & 
\displaystyle{ \frac{1}{16} \left( \frac{202}{3} - \frac{20}{9}\nf
                                                          \right) },
\\
\nl
\gamma_{2}^{(\nf)} & = &  
\displaystyle{ \frac{1}{64} \left\{ 1249 
             - \left[ \frac{2216}{27} + \frac{160}{3} \,\ztri \right] \nf
             - \frac{140}{81} \nfS \right\} }.
\label{betagamma}
\eqa
The renormalisation scale dependence of the coupling constant can be parameterised at 3-loop level 
as function of $\Lambda^{(nf)}_{\rm\overline{MS}} $
\begin{eqnarray}
\label{running_formula}
\alpha_s(\mu) & = & \frac{\pi}{\beta_0\ln(\mu^2/\Lambda^2)}\bigg[
1-\frac{\beta_1}{\beta_0^2}\frac{\ln\left[\ln(\mu^2/\Lambda^2)\right]}
{\ln(\mu^2/\Lambda^2)}
+\frac{1}{\beta_0^2\ln^2(\mu^2/\Lambda^2)}\times
\nonumber \\
&&\hphantom{\frac{\pi}{\beta_0\ln(\mu^2/\Lambda^2)}\bigg[}\! 
 \times \left( \frac{\beta_1^2}{\beta_0^2}\left\{\ln^2
\left(\frac{\mu^2}{\Lambda^2}\right)
-\ln\left[\ln\left(\frac{\mu^2}{\Lambda^2}\right)\right]
-1 \right\} +
\frac{\beta_2}{\beta_0}\right)\bigg] \, .\qquad
\end{eqnarray}

Technically, for a given input value of $\alpha_s(\mz)$, Eq.~\ref{running_formula} is solved 
numerically 
for $\nf=5$ in order to obtain $\Lambda^{(5)}_{\rm\overline{MS}} $. The scale parameters for $\nf=4$ 
and $\nf=3$, $\Lambda^{(4)}_{\rm\overline{MS}} $ and $\Lambda^{(3)}_{\rm\overline{MS}} $, required for 
the evolution of $\alpha_s$ to the scales of the quark pole masses $M_b$ and $M_c$, are derived using 
the matching condition  
\bqa
\ln\left(\frac{\LMSBn}{\LMSBnml}\right)^2&=&\beta^{(\nf-1)}_0\Biggl\{
\left(\beta^{(\nf)}_0-\beta^{(\nf-1)}_0\right)L_{\sss{M}}
+\left(\frac{\beta^{(\nf)}_1}{\beta^{(\nf)}_0}-\frac{\beta^{(\nf-1)}_1}
{\beta^{(\nf-1)}_0}\right)\ln
L_{\sss{M}} 
\nll 
\nll
&&
-\frac{\beta^{(\nf-1)}_1}{\beta^{(\nf-1)}_0}\ln\frac{\beta^{(\nf)}_0}
{\beta^{(\nf-1)}_0}
+\frac{\beta^{(\nf)}_1}{\left(\beta^{(\nf)}_0\right)^2}
 \left(\frac{\beta^{(\nf)}_1}{\beta^{(\nf)}_0}-\frac{\beta^{(\nf-1)}_1}
{\beta^{(\nf-1)}_0}\right) 
 \frac{\ln L_{\sss{M}}}{L_{\sss{M}}}
\nll 
\nll
&&
+\frac{1}{\beta^{(\nf)}_0}\Biggl[\Biggl(\frac{\beta^{(\nf)
    }_1}{\beta^{(\nf)  }_0}\Biggl)^2 
-\Biggl(\frac{\beta^{(\nf-1)}_1}{\beta^{(\nf-1)}_0}\Biggl)^2 
-\frac{\beta^{(\nf)}_2}{\beta^{(\nf)  }_0} 
+\frac{\beta^{(\nf-1)}_2}{\beta^{(\nf-1)}_0} 
-\frac{7}{72}\Biggr]\frac{1}{L_{\sss{M}}}\Biggr\}, 
\nll
\eqa
with
\bqa
L_{\sss{M}} & = & \ln \frac{\MqS}{\LMSBnS}\;.
\eqa
%---

\subsection{Quantifying higher order contributions} 
\label{quantify}
The theoretical prediction for the widths and derived EW observables consists 
of three basic ingredients, each incorporating different classes of 
higher order contributions: the NNLO massless terms, the quark mass corrections 
and the mixed EW$\otimes$QCD corrections. For the evaluation of associated perturbative 
uncertainties it is essential to study the impact and size of the these contributions 
to the full theory. The sensitivity of widths and realistic observables to the 
three classes of corrections is illustrated in Table~\ref{tab:quantify}, where the 
relative change of the full prediction with respect to downgraded calculations 
neglecting certain terms is given.  
\TABULAR[h]{|l|rr|rrr|}{
\hline
Theory          & NNLO &       & NLO   & without quark   & without mixed \\
                &      &      &       &  mass corrections         & corrections \\ \hline
$\Gamma_u$      & $300.057$ & MeV & $-0.13$ & $ 0.00$ & $ 1.66$  \\ 
$\Gamma_{d,s}$  & $382.901$ & MeV & $ 1.28$ & $ 0.00$ & $ 1.45$  \\
$\Gamma_c$      & $299.994$ & MeV & $-0.18$ & $ 0.21$ & $ 1.67$ \\
$\Gamma_b$      & $375.807$ & MeV & $ 1.23$ & $ 4.13$ & $ 1.47$  \\ 
$\Gamma_h$      & $1741.66$ & MeV & $ 0.77$ & $ 0.93$ & $ 1.53$  \\
$\Gamma_l$      & $83.796$  & MeV & $ 0.00$ & $ 0.00$ & $ 1.19$ \\
$\Gamma_Z$      & $2495.08$ & MeV & $ 0.54$ & $ 0.65$ & $ 1.37$\\
$\sigma^0_h$    & $41.4798$ & nb  & $-0.31$ & $-0.37$ & $-0.03$ \\
$\sigma^0_l$     & $2.0002$ & nb  & $-1.08$ & $-1.29$ & $-0.37$ \\
$R_Z$               & $20.737$ &  & $ 0.78$ & $ 0.93$ & $ 0.34$ \\ \hline
                & nominal value & & \multicolumn{3}{|c|}{relative change [$\permil$]} \\ \hline
}
{\label{tab:quantify} Nominal values of widths and EW observables and their 
relative change in per mil observed when certain classes of corrections are omitted.  
The calculations are carried out for $\alpha_s(\mz)=0.1185$.}
The change from NNLO to NLO for the quarkonic widths is between one per mil for down-type quarks 
and 0.1 per mil for up-type quarks, this large difference is generated by flavour-dependent contributions to   
$R^{\fq}_{\sss{A}}$ and $\gZq$. Neither NNLO nor quark mass corrections have any sizeable 
impact on the leptonic width, but the mixed EW$\otimes$QCD corrections entail changes 
of 1.5 per mil to all widths, larger than the final state NNLO corrections. Not 
surprisingly, the quark mass corrections are as large as 4 per mil for b-quarks 
but drop to 0.2 per mil for c-quarks. The total width absorbs an average 
effect induced by the widths and the other realistic observables are 
in general less sensitive to these corrections cancelling in the ratio of widths. Among the 
realistic observables $\sigma^0_h$ is least and $\sigma^0_l$ most sensitive to 
the higher order corrections, their size ranges between $0.3$ and $1.3\permil$.

%%%%%%%
\section{Theoretical uncertainties for EW observables}\label{sec:uncert}
%%%%%%%
The sensitivity of a given electroweak observable to $\alpha_s$ 
originate on one side from the QCD corrections incorporated in the radiator functions 
and on the other side from the mixed EW$\otimes$QCD corrections to the effective couplings. 
A measurement of $\alpha_s$ using EW observables is subject to a systematic uncertainty 
stemming from missing higher orders in the perturbation series. 
The yet uncalculated higher orders 
are inherently difficult to access. A conventional method of estimating the perturbative 
uncertainty consists of a variation of the renormalisation scale $\mu$. The 
natural scale of the process is usually taken to be $\sqrt{s}$ in $\epem$ annihilation,
 and subsequently in the case of 
Z peak  
observables $\mu$ is set to $\mz$. Neither this particular choice for 
the nominal scale nor the range of variation for $\mu$ are unambiguous \cite{Hasko}. 
Following the convention applied in 
analyses of $\epem$ event-shape variables, the perturbative uncertainty is 
estimated by changing $\xmu=\mu/\sqrt{s}$ in the range $1/2 \leq \xmu \leq 2$.

A variation of 
the renormalisation scale induces a change of the value of $\alpha_s(\mu)$ as given in Eq.~\ref{running_formula}. 
At NLO and beyond this change is compensated by a modification of the (N)NLO terms, resulting in a 
residual dependence at (N)NNLO. The details of the scale dependence of the radiator functions 
is discussed below. 

The mixed $\cal O$$(\alpha\alpha_s)$ corrections are complete only at leading 
order in $\alpha_s$, but the dominant three-loop correction in Eq.~\ref{leading} leading 
in $m_t^2$ is included. For these $\cal O$$(\alpha\alpha_s^2)$ terms the explicit scale 
dependence is taken into account.  

Given the overall small size of the EW$\otimes$QCD corrections,  
they do not contribute significantly to the scale dependence of the realistic observables.    

%In the present case of a complete NNLO calculation for the radiator functions, another 
%estimate of missing higher orders can be deduced from the difference between 
%the NNLO and the NLO result. 

%\subsection{NNLO-NLO difference}
%The difference between the full NNLO 
%result and the truncation at NLO can be used to assess the theoretical systematic uncertainty. Provided a 
%fast convergence of the perturbation series, this difference can be regarded as a rough estimate 
%of the 'true` error which would be the difference between the NNLO result and the complete theory. In fact this 
%difference, containing the incomplete NLO result, should be seen as an uncertainty estimate of measurement 
%based on the NLO theory, while an improved NNLO-based determination should have a smaller uncertainty. 
%Technically, the reduction to NLO is straightforward, it is sufficient to drop the terms in
% Eqs.\ref{rvfact}-\ref{rafact} proportional to $\alpha_s^3$.

\subsection{Renormalisation scale dependence}
Dimensional regularisation introduces a renormalisation scale $\mu$ at which the coupling constant is defined. 
Thereby, the coefficients in the expansion of $R_{V,A}^q$ acquire an explicit dependence on this scale, 
which is only at all orders completely compensated  by the scale dependence of $\alpha_s(\mu)$. For a NNLO 
calculation, the residual scale dependence is N$^3$LO. The nominal value 
of the $\mu$ scale is set to the scale of the process $\mu^2=s$. 

The expression for $R_{\sss V,A}^q$ in Eqs.~\ref{rvfact}, \ref{rafact} are valid only for $\mu^2=s$, for different 
renormalisation scales terms proportional to powers of $\ln \mu^2/s$ appear. For a generic power series of the type 
\begin{equation}
R = \sum_{i=0}^n c_n\left(\frac{\alpha_s}{\pi}\right)^n\;,
\end{equation}
the NLO coefficient $c_2$ becomes a function of $\mu$
\begin{equation}
c_2 \rightarrow c_2(\mu)=c_2+\beta_0 c_1 \ln \frac{\mu^2}{s}\;.
\end{equation}
It is important to note that two quantities depend on the renormalisation scale in Eqs.~\ref{rvfact} and \ref{rafact}: 
the coupling constant $\alpha_s(\mu)$ (Eq.~\ref{running_formula}) and the running masses $m_q(\mu)$ 
(Eq.~\ref{runmasc}).    

In order to to simplify the formulae for the scale dependence of the radiator functions, it is convenient to 
re-order Eqs.~\ref{rvfact} and \ref{rafact} in terms of powers of the running masses:
\begin{equation}
R_{V,A}^q(\mu)  =  \sum_{f=q,q^{\prime},c,b}\sum_{i=0}^3 \frac{m_f^{2i}(\mu)}{s^i} \sum_{j=0}^3d_{ij,f}^{A,V}\asb^j \;,
\label{masum}
\end{equation}  
where $\asb = \frac{\alpha_s(\mu^2)}{\pi}$. For each quark not only the mass 
of the actual quark, but also the mass  the b- and c-quark masses intervene 
in the radiator functions, each with different coefficients. 
In the $\overline{\rm MS}$ scheme the expansion of 
the scale evolution of powers of 
the running masses reads as
\begin{eqnarray}
m^2_q(s) & = & m^2_q(\mu)\Biggl(1 + 2 \gamma_0 L \asb + \left(\gamma_0 \beta_0 L^2 + 2 \gamma_1 L + 
2 \gamma_0^2 L^2\right) \asb^2 \\ \label{eq:m2}
 & & + \biggl[\frac{2}{3}\gamma_0\beta_0^2 L^3 + \gamma_0\beta_0 L^2 + 2 \gamma_2 L + 
2 \gamma_0^2\beta_0 L^3 + 4\gamma_0 \gamma_1 L^2 + \frac{4}{3} \gamma_0^3 L^3 \biggr]\asb^3\Biggr)\; , \nonumber \\ 
m^4_q(s) & = & m^4_q(\mu)\Biggl(1 + 4\gamma_0  L \asb + \left(2\gamma_0\beta_0 L^2 + 4\gamma_1 L 
+ 8 \gamma_0^2 L^2\right)\asb^2 \Biggr] \; , \\\label{eq:m4}
m^6_q(s) & = & m^4_q(\mu)\left(1 + 6\gamma_0 L \asb \right)\; ,\label{eq:m6}
\end{eqnarray} 
with $L=\ln x_\mu^2$. The new coefficients $d_{i,j}^{A,V}$ 
are related to those of Eqs.~\ref{rvfact} and \ref{rafact} by the following formulae.\\ 

\noindent Massless terms:
\begin{eqnarray}
d_{00,q}^{V} = 1 + \frac{3}{4}Q_q^2\frac{\alpha(s)}{\pi} & , &  
d_{00,q}^{A} = d_{00,q}^{V},  \\
d_{10,q}^{V} = 1 - \frac{1}{4}Q_q^2\frac{\alpha(s)}{\pi} & , &  
d_{10,q}^{A} = d_{10,q}^{V}, \\
d_{20,q}^{V} = C_{02} + C_2^t\left(\frac{s}{m_t^2}\right) & , & 
d_{20,q}^{A} = C_{02} + \left(C_2^t - 2I_q^{(3)} {\cal I}^{(2)}\right)
\left(\frac{s}{m_t^2}\right), \\
d_{30,q}^{V} = C_{03} & , & 
d_{30,q}^{A} = C_{03} - 2I_q^{(3)} {\cal I}^{(2)}; \\ 
\end{eqnarray}
%---
 
\noindent Terms in $m_q^2$:
\begin{eqnarray}
d_{02,q}^{V} = 0 & , &
d_{02,q}^{A} = C_{20}^{A}, \\
d_{12,q}^{V} = C_{21}^{V} & , &
d_{12,q}^{A} = C_{21}^{A}, \\
d_{22,q}^{V} = C_{22}^{V} & , &
d_{22,q}^{A} = C_{22}^{A} + 6\left(3 + \ln \frac{m_t^2}{s}\right)
-10\frac{s}{m_t^2}\left(\frac{8}{81}+\frac{1}{54}\ln\frac{m_t^2}{s}\right), \\
d_{32,q}^{V} = C_{23}^{V} & , &
d_{32,q}^{A} = C_{23}^{A}, \\
d_{32,b}^{V} = d_{32,c}^{V} = C_{23} & , &
d_{32,b}^{A} = d_{32,c}^{A} = C_{23}; \\
\end{eqnarray} 
%--- 

\noindent Terms in $m_q^4$:
\begin{eqnarray}
d_{04,q}^{V} = 0 & , &
d_{04,q}^{A} = C_{40}^{A}, \\
d_{14,q}^{V} = C_{41}^V & , &
d_{14,q}^{A} = C_{41}^{A}, \\
d_{24,q}^{V} = C_{42}^V + C_{42}^{V,L} \ln \frac{m_q^2(s)}{s} & , &
d_{24,q}^{A} = C_{42}^{A} + C_{42}^{A,L} \ln \frac{m_q^2(s)}{s}, \\
d_{24,b}^{V} = d_{24,b}^{A} =   C_{42} - \ln \frac{m_b^2(s)}{s} & , &
d_{24,c}^{V} = d_{24,c}^{A} =   C_{42} - \ln \frac{m_c^2(s)}{s}\\
d_{24,\hat q}^{V} = d_{24,\hat q}^{A} = 12; &  & \\
\end{eqnarray} 
%---

\noindent Terms in $m_q^6$:
\begin{eqnarray}
d_{06,q}^{V} = 8 & , & d_{06,q}^{A} = 0, \\
d_{16,q}^{V} = \frac{16}{27}\left(155 + 6 \ln\ln \frac{m_q^2(s)}{s}\right) & , & d_{16,q}^{A} = 0. \\
\end{eqnarray} 
Finally, to get the renormalisation scale dependence of the radiator functions, the terms of the 
type $m_q^j(s) \sum_i d_{ij}\alpha_s^i(s)/\pi$ in Eq.~\ref{masum}, dropping for clarity the 
axial-vector/vector and 
flavour indices, have to be replaced by the following expressions:
\begin{eqnarray} 
\sum_{j=0}^{3} d_{0j}\frac{\alpha_s^j(s)}{\pi} & \rightarrow & d_{00} + d_{10}\asb + \left(d_{20} + d_{10}\beta_0 L \right)\asb^2 \\
& & + \left(d_{30} + \left(d_{10}\beta_1 + 2d_{20}\beta_0\right) + d_{10}\beta_0^2 L^2\right)\asb^3 , \nonumber \\
m_q^2(s)\sum_{j=0}^{3} d_{j2}\frac{\alpha_s^j(s)}{\pi} & \rightarrow & 
m_q^2(\mu)\Biggl[d_{02}+\left(d_{12}+2d_{02}\gamma_0 L \right)\asb \\
& & +\left(d_{12}\beta_0L + d_{22} + 2d_{12}\gamma_0 L +d_{02}\gamma_0\beta_0L^2 + 2d_{02}\gamma_1L + 2d_{02}\gamma_0^2L^2\right)\asb^2 \nonumber \\
& & +\Biggl(d_{12}\beta_0^2L^2 + d_{12}\beta_1L + d_{32} + 2d_{22}\beta_0L +\frac{2}{3}d_{02}\gamma_0\beta_0^2L^3
+ d_{02}\gamma_0\beta_1L^2 \nonumber \\
& & + 2d_{02}\gamma_1\beta_0L^2+2d_{02}\gamma_2L + 2d_{02}\gamma_0^2\beta_0L^3 + 4d_{02}\gamma_0\gamma_1L^2 
+\frac{4}{3}d_{02}\gamma_0^3L^3 \nonumber \\
& & + 3d_{12}\gamma_0\beta_0L^2 + 2d_{22}\gamma_0L + 2d_{12}\gamma_1L +2d_{12}\gamma_0^2L^2 \Biggr)\asb^3\Biggr], \nonumber \\
m_q^4(s)\sum_{j=0}^{2} d_{j4}\frac{\alpha_s^j(s)}{\pi} & \rightarrow & 
m_q^4(\mu)\Biggl[d_{04} + \left(d_{14}+4d_{04}\gamma_0L\right)\asb \\
& & \left(d_{14}\beta_0L + d_{24} + 2d_{04}\gamma_0\beta_0L^2 + 4d_{04}\gamma_1L + 8 d_{04} \gamma_0^2L^2+ 4d_{14}\gamma_0L\right)\asb^2\Biggr] \nonumber \\
m_q^6(s)\sum_{j=0}^{1} d_{j6}\frac{\alpha_s^j(s)}{\pi} & \rightarrow & 
m_q^6(\mu)\Biggl[d_{06} + \left(d_{16}+6d_{06}\gamma_0L\right)\asb\Biggr]
\end{eqnarray}  
%---

The dependence of the radiator functions on the logarithm of the renormalisation scale $\ln x_\mu$ 
for a fixed input value of $\alpha_s(\mz)$ is shown for each quark flavour  
in Fig.~\ref{fig:Rasmurarv}. 
\setlength{\unitlength}{1mm}
\FIGURE[h!]{\epsfig{file=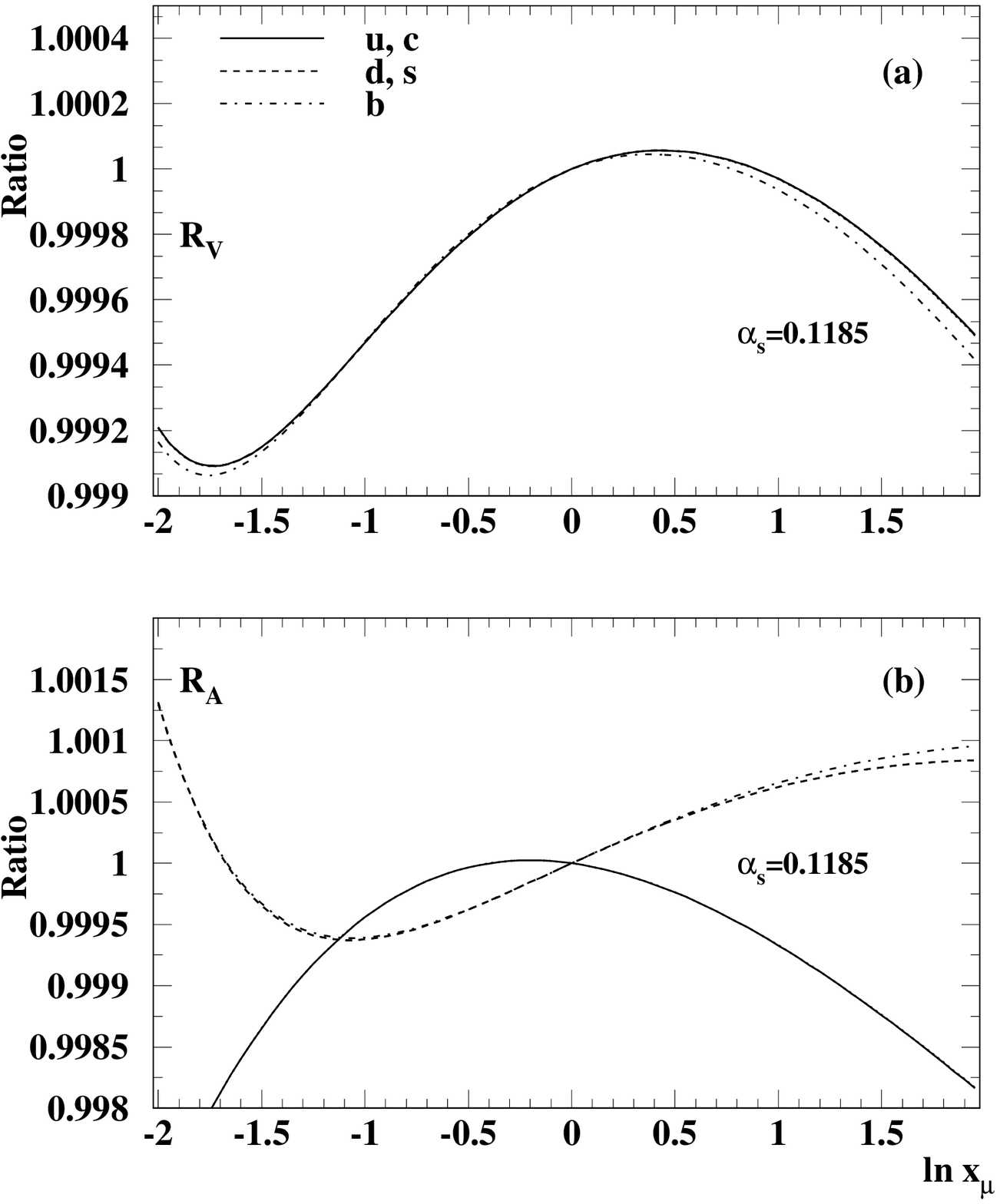, width=0.8\textwidth}
\caption{ Dependence of the radiator functions on the renormalisation scale.  
The vector correction $R^q_{\sss{V}}$ (a) and the 
axial vector correction $R^q_{\sss{A}}$ (b) are shown for the quarks $q=u,d,s,c,b$ as function 
of $\ln x_\mu$, normalised to their value at $x_\mu=1$.\label{fig:Rasmurarv}}}

The shape of the scale dependence of $R^q_{\sss{V}}$ is almost identical for all flavours, 
except a small quark mass modification for the b-quark. A maximum in $R^q_{\sss{V}}$ 
appears around $\ln\xmu=0.5$ and a minimum at $-1.75$, spanning a difference of one 
per mil. The overall scale 
dependence of the axial-vector component $R^q_{\sss{A}}$ is twice as large as 
the one of $R_V^q$. As already observed for the dependence on $\alpha_s$ in 
Fig.~\ref{fig:as_rarv}, the shape of $R^q_{\sss{A}}$ is clearly different for up- and 
down-type quarks. For up-type quarks $R^q_{\sss{A}}$ becomes maximal at $\ln\xmu=-0.3$, 
the maximum is close to the nominal value at $\ln\xmu=0$. The shape of the scale dependence 
is opposite for down-type quarks: a minimum appears at $\ln\xmu=-1.1$. 
Considering the range of variation for $x_\mu$ from $1/2$ to 
$2$, corresponding to a range for $\ln x_\mu$ from $-0.7$ to $0.7$, it appears 
that largest deviation from the nominal point at $\ln \xmu =0$ is not always 
obtained at the endpoints, but sometimes at smaller variations. Therefore, when assessing 
the uncertainty for the observables 
studied in the following, the endpoints of $\ln x_\mu$ ($\ln \xmu^-$ and $\ln \xmu^+$) 
have been chosen to 
correspond to the largest change in the observables, within the pre-defined range 
$|\ln x_\mu|<0.7$. 

The dependence of the effective couplings on $\ln x_\mu$ is shown in Fig.~\ref{fig:Rasmuroge}.    
\setlength{\unitlength}{1mm}
\FIGURE[h!]{\epsfig{file=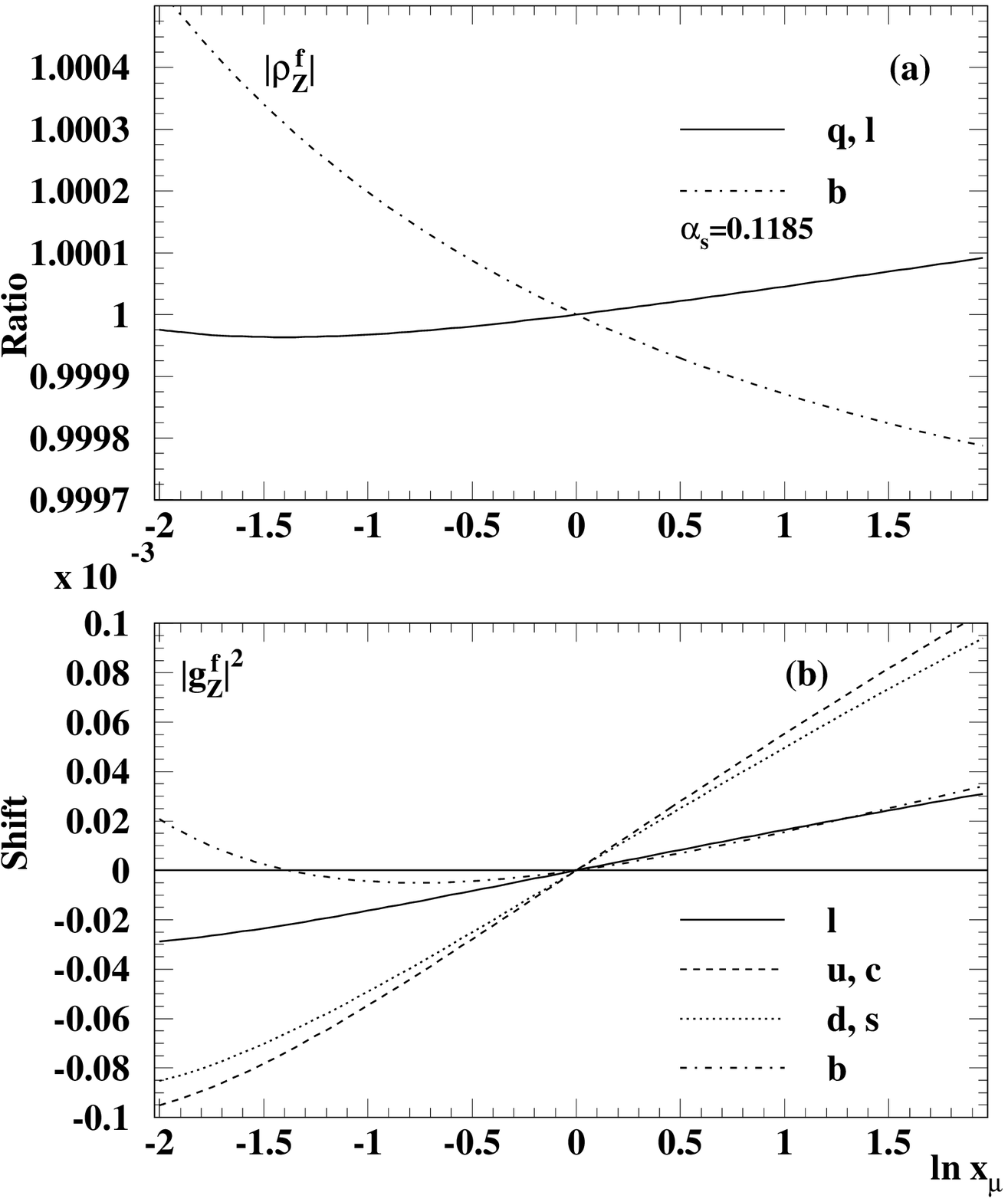, width=0.8\textwidth}
\caption{Dependence of the effective couplings $|\rZf|$ (a) 
and $|\gZf|^2$ (b) on the renormalisation scale.\label{fig:Rasmuroge}}}
The quantities $|\rZf|$ and $|\gZf|^2$, as they 
appear in the expressions for the widths, are shown as ratio for $\rZf$ but 
as shift for $\gZf$, i.e. $|\gZf|^2(\ln\xmu) -|\gZf|^2(\ln\xmu=0)$. 
The shape of the couplings scale dependence 
is rather different from the radiator functions dependence and less structured, 
given by the interference of complete two-loop and incomplete leading three-loop 
corrections to the mixed corrections. 
For all light quarks and leptons the relative change of $|\rZf|$ is only about 
$10^{-4}$, for the b-quark much larger and amounts to $0.8$ per mil.
The absolute change of $|\gZf|^2$ is about $2 \cdot 10^{-4}$ for light quarks and  $0.6 \cdot 10^{-4}$ 
for leptons and slightly less for b-quarks, corresponding to a relative change of less than 
two per mil for all quarks but almost one percent for the leptons. 

Turning to the widths, for hadronic final states both the effective couplings and 
the radiator functions discussed 
above contribute to scale dependence. For the leptonic widths only the effective couplings 
depend on the renormalisation scale through the mixed corrections. The 
relative magnitude of contributions is given by the formulae for the widths, 
Eq.~\ref{lwidth} for the leptons and Eq.~\ref{defzwidthq} for the quarks. 
The scale dependence of the widths and of the realistic observables are shown 
in Fig.~\ref{fig:rasmu_wi}. 
 \setlength{\unitlength}{1mm}
\FIGURE[h!]
{\epsfig{file=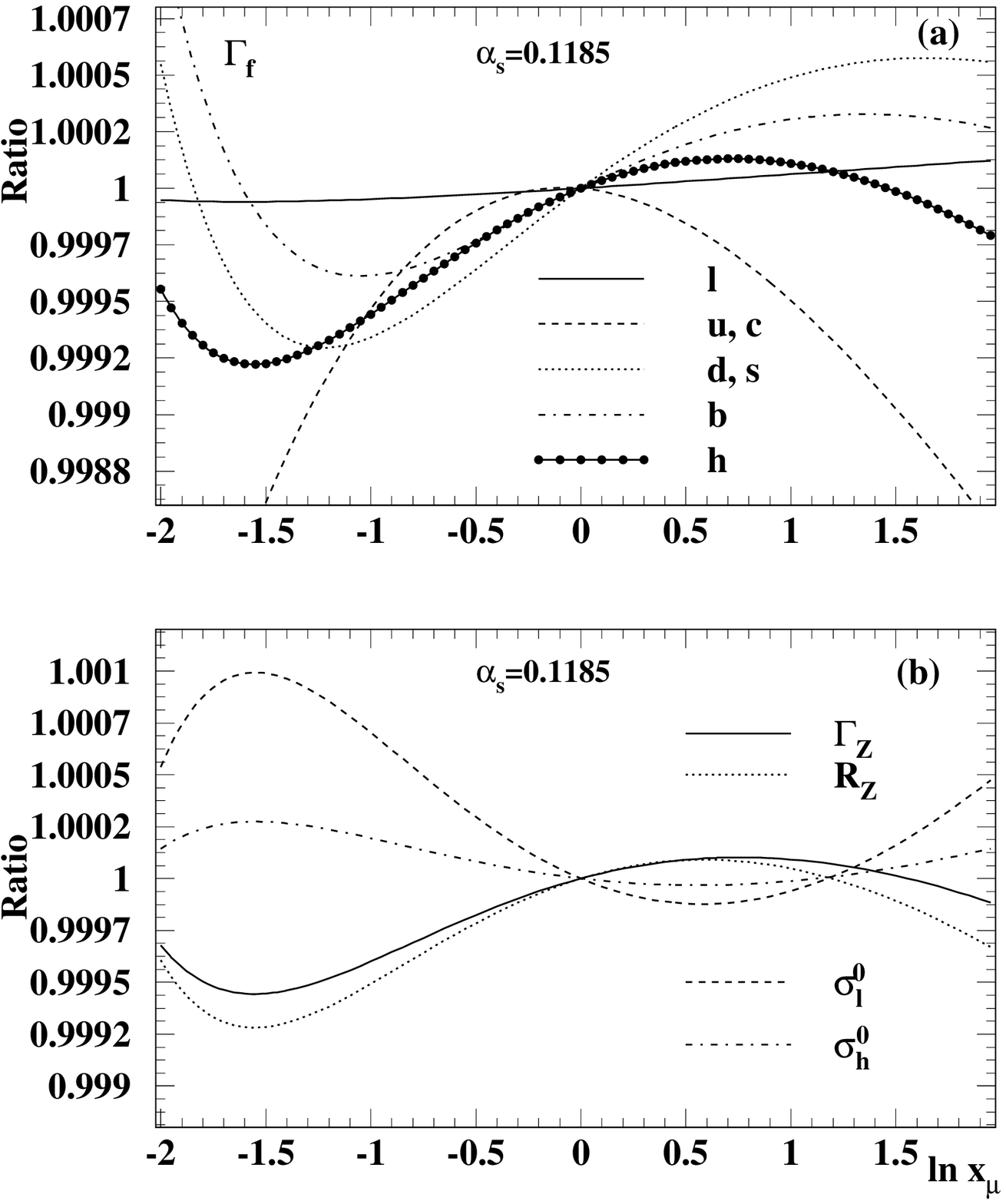 , width=0.8\textwidth}
\caption{Dependence of the widths (a) and selected 
realistic observables (b) on the renormalisation scale for $\alpha_s(\mz)=0.1185$.\label{fig:rasmu_wi}}}

The shape of the scale dependence for the widths depends on the fermion type. 
For the u- and c-quark a maximum is found close to $\ln \xmu=0$, and variations 
of $\ln \xmu$ in any direction entail a decrease of the width. The partial widths 
into d- and s-quarks increase monotonically from $\ln \xmu=-0.7$ to $\ln \xmu=0.7$, similarly  
for the width of the b-quark, albeit with a flatter shape. The total hadronic widths 
of the Z boson emerges as sum of the quarkonic contributions, leading to an 
average shape of its scale dependence with a minimum at $-1.6$, a maximum at $0.7$ with 
a difference of one per mil between them.  

The scale dependence of the leptonic widths is 
clearly much weaker than that of the quarkonic counterparts. In the central 
range of $\ln \xmu$ the leptonic width changes by about a tenth of the hadronic width's change. 
The sensitivity of $\Gamma_l$ to the renormalisation scale arise through the EW$\times$QCD 
corrections, essentially those incorporated in $\rZl$.

\subsection{Perturbative uncertainties for observables}
The perturbative uncertainty for widths and realistic observables are 
defined as the difference between the observables value at $\xmu=1$ and 
at $\xmu^{+/-}$, for a given input value of $\alpha_s(\mz)$. The range of 
variation for $\xmu$, defined by $\xmu^+$ and $\xmu^-$, is chosen to generate 
the maximum and minimum values of the observables within the pre-defined 
range $1/2 \leq \xmu \leq 2$. The absolute size of the perturbative uncertainty 
depends on the input value of $\alpha_s$, for an observable calculated at NNLO 
its uncertainty is scaling with $\alpha_s^4$. The relative systematic uncertainty 
for the widths and observables is shown in Fig.~\ref{fig:obserr} as function of $\alpha_s(\mz)$. 
Here and in Table~\ref{tab:obserr} the relative uncertainties for a generic observable $O$ 
are defined by 
\begin{equation}
\Delta O = \frac{\left[O(\xmu^{\pm})-O(\xmu=1)\right]}{O(\xmu=1)} \; .
\end{equation} 

\setlength{\unitlength}{1mm}
\FIGURE[h!]{\epsfig{file=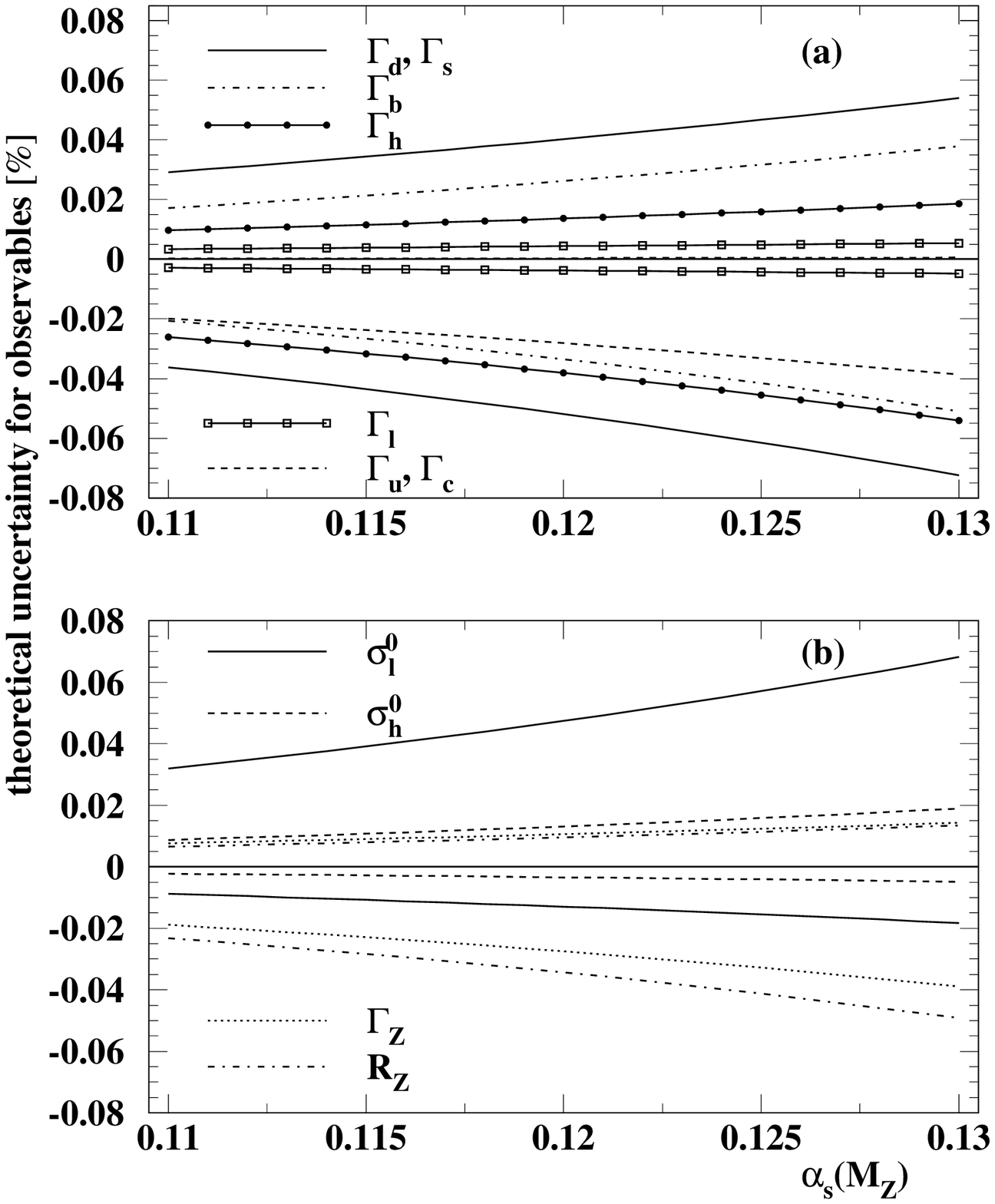, width=0.8\textwidth}
\caption{Relative perturbative uncertainty in percent for the widths (a) and selected 
realistic observables (b) as a function of $\alpha_s(\mz)$.\label{fig:obserr}}}

The positive and negative systematic uncertainties are generally asymmetric, for 
the widths of the u- and c-quark only the negative uncertainty contributes, given 
the particular shape of their scale dependence. The uncertainty for the width of 
down-type quarks is typically $\pm 0.05 \%$, for up-type quarks it is $-0.03 \%$ 
and for the leptonic width a factor of 10 smaller $\pm 0.005 \%$. Among the 
realistic observables $\sigma^0_l$ has the largest uncertainty of about $+0.05\%$, 
the other variables uncertainty is about half that size. 

For selected values of  $\alpha_s(\mz)$ the uncertainties for the observables 
are given in Table~\ref{tab:obserr}, where the scale variation endpoints 
$\xmu^{\pm}$ are listed in the last column.

\TABULAR[h]{|c|rrrrr|c|}{
\hline
$\alpha_s(\mz)$ &  $0.110$  & $0.115$   & $0.120$   & $0.125$   & $0.130$ & $\xmu^{\pm}$\\ \hline
$\Gamma_u$      & $0.002$ & $0.003$ & $0.003$ & $0.004$ & $0.006$ & $0.925$\\ 
                & $-0.199$ & $-0.238$ & $-0.282$ & $-0.331$ & $-0.387$ & $2.0$\\ \hline 
$\Gamma_{d,s}$  & $0.292$ & $0.344$ & $0.403$ & $0.468$ & $0.540$ &  $2.0$ \\
                & $-.363$ & $-.435$ & $-.519$ & $-.615$ & $-.725$ & $0.5$ \\ \hline
$\Gamma_c$      & $0.002$ & $0.003$ & $0.003$ & $0.004$ & $0.006$ & $0.925$ \\
                &$-.198$ & $-.237$ & $-.281$ & $-.331$ & $-.387$ & $2.0$ \\ \hline
$\Gamma_b$      & $0.172$ & $0.214$ & $0.262$ &  $0.317$ & $0.379$ & $2.0$ \\ 
                & $-.207$ & $-.266$ & $-.335$ & $-.416$ & $-.510$ & $0.5$ \\ \hline
$\Gamma_h$      & $0.098$ & $0.116$ & $0.137$ & $0.160$ & $0.186$ & $2.0$ \\
                & $-.261$ & $-.317$ & $-.381$ & $-.455$ & $-.540$ & $0.5$\\ \hline
$\Gamma_l$      & $0.035$ & $0.039$ & $0.044$ & $0.049$ & $0.054$ & $2.0$\\
                & $-.029$ & $-.033$ & $-.038$ & $-.043$ & $-.048$ & $0.5$\\ \hline
$\Gamma_Z$      & $0.076$ & $0.090$ & $0.106$ & $0.124$ & $0.143$ & $2.0$\\
                & $-.189$ & $-.229$ & $-.275$ & $-.328$ & $-.389$ & $0.5$\\ \hline 
$\sigma^0_h$    & $0.088$ & $0.108$ & $0.131$ & $0.158$ & $0.190$ & $0.5$\\
                & $-.023$ & $-.028$ & $-.034$ & $-.041$ & $-.049$ & $1.75$\\ \hline
$\sigma^0_l$    & $0.320$ & $0.391$ & $0.475$ & $0.571$ & $0.682$ & $0.5$ \\
                & $-.088$ & $-.107$ & $-.130$ & $-.155$ & $-.183$ & $1.8$ \\ \hline
$R_Z$           & $0.066$ & $0.080$ & $0.096$ & $0.114$ & $0.135$ & $1.8$ \\
                & $-.232$ & $-.284$ & $-.343$ & $-.413$ & $-.492$ & $0.5$ \\ \hline
}
{\label{tab:obserr}
Systematic perturbative uncertainties in per mil of EW observables for 
different values of $\alpha_s(\mz)$. The last colum indicates the values 
of the renormalisation scale $\xmu^{\pm}$ corresponding to the maximum variation of the observables with 
respect to their nominal values at $\xmu=1$, within the pre-defined variation range 
$1/2 \leq \xmu \leq 2$}%%%%%%%%%%%%%%%%%
\subsection{Contributions to the scale dependence}
\label{quantifyscale}
The three classes of higher order corrections analysed in Section~\ref{quantify} 
contribute to the perturbative uncertainty of the observables via their 
different evolution under the renormalisation scale variation. The size 
of the NNLO, quark mass and mixed EW$\otimes$QCD corrections, given in Table~\ref{tab:quantify}, 
are typically at the level of one per mil, while the perturbative uncertainties 
are about 0.1 per mil (see Table~\ref{tab:obserr}). The absolute size of the corrections 
alone is not a reliable indicative of their contribution to the uncertainty, 
which must be evaluated from their scale dependence. 

The scale uncertainty is therfore re-calculated without the NNLO 
contribution (i.e. dropping terms in $\cal O$$(\alpha_s^3)$ in Eqs.~\ref{rvfact}, \ref{rafact}),   
in which case also the evolution of $\alpha_s(\mu)$ in Eq.~\ref{running_formula} has to 
be performed at NLO. 

The contribution from the quark mass corrections is estimated by switching off the explicit 
scale dependence of the running quark masses (i.e. setting $m_q^p(s)=m_q^p(\mu)$ in 
Eqs.~\ref{eq:m2} - \ref{eq:m6}). 

The impact of scale dependence of the mixed EW$\otimes$QCD corrections is tested by  
eliminating the terms in the $\cal O$$(\alpha\alpha_s^2)$ corrections depending explicitly 
on $\xmu$. 

The scale dependence obtained under these conditions is compared to the nominal scale 
dependence of the complete NNLO prediction in Fig.~\ref{fig:zxmu_obs}, considering as 
example $\Gamma_Z$ and $\sigma^0_h$. 
\setlength{\unitlength}{1mm}
\FIGURE[t!]
{\epsfig{file=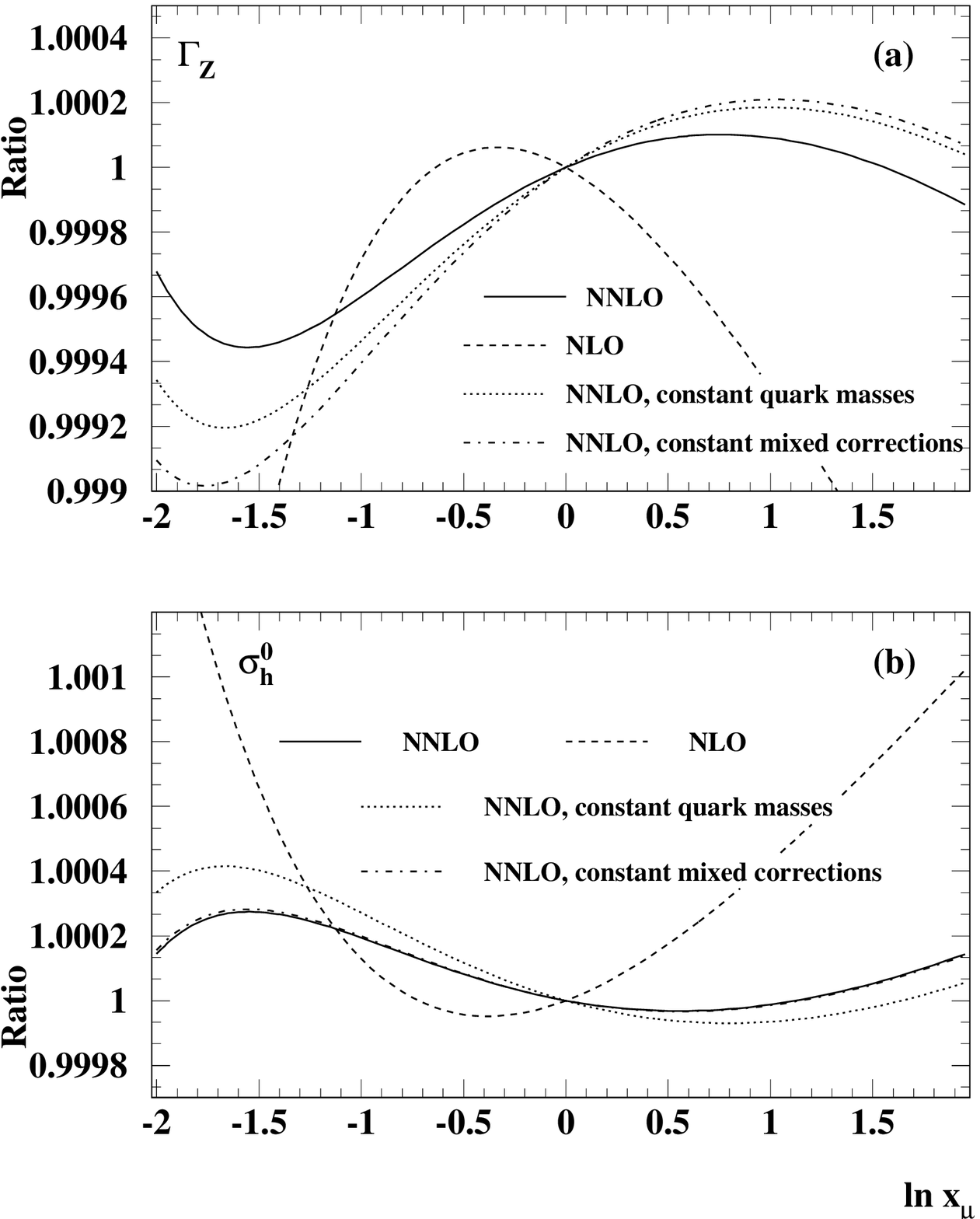, width=0.8\textwidth}
\caption{Renormalisation scale dependence 
of $\Gamma_Z$ (a) and $\sigma^0_h$ (b) using the full NNLO prediction  
compared to reduced theories at NLO only, neglecting the scale dependence 
of the running quark masses and of the mixed EW$\otimes$QCD corrections,    
for $\alpha_s(\mz)=0.1185$.\label{fig:zxmu_obs}}}

The change from NNLO to NLO entails a strongly increased scale 
dependence for the quarkonic widths, but with opposite effect on 
up- and down-type quarks, resulting in a weaker enhancement for 
the realistic observables. The uncertainties for the observables 
obtained as before by a scale variation between $\xmu^{+}$ and 
$\xmu^{-}$ are summarised in Table~\ref{tab:mucontri} for the 
theories with modified scale dependence of higher order corrections. 
For the realistic observables the downgrade from NNLO to NLO 
results in an increase by a factor of two of the perturbative 
uncertainty and the asymmetry between positive and negative 
uncertainty is also enlarged.      
\TABULAR[h]{|c|rrrr|}{
\hline
Theory          & NNLO        & NLO   & fixed quark   & fixed mixed \\
                &             &       &  mass corrections         & corrections \\ \hline
$\Gamma_u$      & $0.003$  & $1.430$  & $0.003$  & $0.004$  \\ 
                & $-0.268$ & $-1.673$ & $-0.268$ & $-0.374$  \\ 
$\Gamma_{d,s}$  & $0.384$  & $0.067$  & $0.384$  & $0.477$  \\
                & $-0.493$ & $-0.851$ & $-0.493$ & $-0.628$  \\
$\Gamma_c$      & $0.003$  & $1.431$  & $0.004$  & $0.004$ \\
                & $-0.268$ & $-1.671$ & $-0.291$ & $-0.374$ \\
$\Gamma_b$      & $0.247$  & $0.0230$ & $0.684$  & $0.339$  \\ 
                & $-0.313$ & $-0.673$ & $-0.879$ & $-0.448$  \\ 
$\Gamma_h$      & $0.130$  & $0.096$  & $0.228$  & $0.226$  \\
                & $-0.361$ & $-0.615$ & $-0.488$ & $-0.502$  \\
$\Gamma_l$      & $0.042$  & $0.042$  & $0.042$  & $0.121$ \\
                & $-0.036$ & $-0.036$ & $-0.036$ & $-0.151$ \\
$\Gamma_Z$      & $0.101$  & $0.063$  & $0.170$  & $0.189$\\
                & $-0.260$ & $-0.419$ & $-0.349$ & $-0.390$\\
$\sigma^0_h$    & $0.124$  & $0.265$  & $0.174$  & $0.126$ \\
                & $-0.032$ & $-0.048$ & $-0.069$ & $-0.033$ \\
$\sigma^0_l$    & $0.448$  & $0.923$  & $0.626$  & $0.477$ \\
                & $-0.122$ & $-0.164$ & $-0.255$ & $-0.140$ \\
$R_Z$           & $0.091$  & $0.115$  & $0.186$  & $0.106$ \\ 
                & $-0.325$ & $-0.657$ & $-0.452$ & $-0.350$ \\ \hline
}
{\label{tab:mucontri}
Perturbative uncertainties (in per mil) for selected observables obtained 
from renormalisation scale variation at $\alpha_s(\mz)=0.1185$. The nominal NNLO uncertainty
 is compared to reduced theories where the scale dependence for certain classes of 
corrections is switched off.}
For realistic observables the shape of the scale dependence without the running 
masses is similar but steeper than in the NNLO case. It has a large impact only 
for the b-quark, where the uncertainty is increased by a factor of three, which 
propagates to an increase of about 50$\%$ for the realistic observables.  

The effect of dropping the scale dependence  
of the mixed corrections has in contrast to the other variations a net impact 
on the leptonic width, since this is the only source of sensitivity to QCD effects. 
The perturbative uncertainty for $\Gamma_l$ is increased by a factor of three, 
the quarkonic width uncertainty by about 30-40 $\%$. The realistic 
observables of ratios of widths are less sensitive to the mixed corrections and 
a moderate enhancement of the uncertainty of a few percent is observed, while 
for $\Gamma_Z$ the effect from quarkonic and leptonic widths adds up to an almost 
doubled uncertainty.

In conclusion the stability of the predictions under scale variations 
depends crucially on the NNLO corrections, 
but depending on the observable also quark mass and to lesser extent mixed corrections 
contribute significantly to the accuracy of the calculations.

%%%%%%%%%%%%%%%%%%
\section{Perturbative uncertainties for $\alpha_s$}\label{sec:aserror}
%%%%%%%%%%%%%%%
In the context of global analyses of world electroweak data \cite{EWG} $\alpha_s(\mz)$ is 
fit together with four other free parameters of the standard model: $M_{\rm H}$, $\mz$, $m_t$ and 
$\Delta\alpha_{had}^{(5)}$. The correlation between $\alpha_s$ and the other parameters is 
small. The observables 
included in the fit with a sizeable sensitivity to $\alpha_s$ are $R_Z$, $\Gamma_Z$ and 
$\sigma^0_h$. The information from the leptonic pole cross section $\sigma^0_l$ 
is included in the other observables. This particular selection of observables 
is the result of an optimisation for the best accuracy for the five free SM parameters, 
which may not necessarily be the optimal for $\alpha_s$ alone when the other parameters 
are fixed to their SM values.    
For example 
$\sigma^0_l$ has actually the best sensitivity 
through the inverse squared radiator functions and   
may be used alone to determine $\alpha_s$ from a single parameter fit, 
avoiding thereby the otherwise required correlations between the observables, and    
the dependence of $\alpha_s$ on the Higgs or top mass may be investigated. 

Having in mind this scenario, the procedure is to determine in a first step 
the perturbative uncertainty for measurements of $\alpha_s$ using single selected 
observables and to estimate in a second step the uncertainty for a global 
fit including several variables. 

The basic principle for the uncertainty estimation was developed in \cite{Hasko}: 
the systematic uncertainty for a given observable and fixed value 
of $\alpha_s(\mz)$ (i.e. as obtained from a fit to this observable) is evaluated in the theory 
by variation of the renormalisation scale $\xmu^- < \xmu < \xmu^+$. The change of the observables 
under the scale variation can also be generated by a variation of the input value 
of $\alpha_s$ at fixed $\xmu=1$. This procedure leads in general to two alternative 
values of $\alpha_s$ corresponding to the changes of the observable for $\xmu^-$ 
and $\xmu^+$. The difference between the nominal value of $\alpha_s$ and these two 
alternatives finally determine the perturbative uncertainty of $\alpha_s$. 

The uncertainty itself depends on the type of observable and on the input 
value of $\alpha_s$. At NNLO the size of the perturbative uncertainty 
scales as $\alpha_s^4$. It is instructive to consider the uncertainty 
for $\alpha_s$ for the pseudo-observable widths in a first step, in order to 
understand their contribution to realistic observables. The systematic uncertainties 
are shown in Fig.~\ref{fig:Zaswi} as function of the input value of $\alpha_s$ 
in a relevant range from $0.11$ to $0.13$.              
\FIGURE[h!]
{\epsfig{file=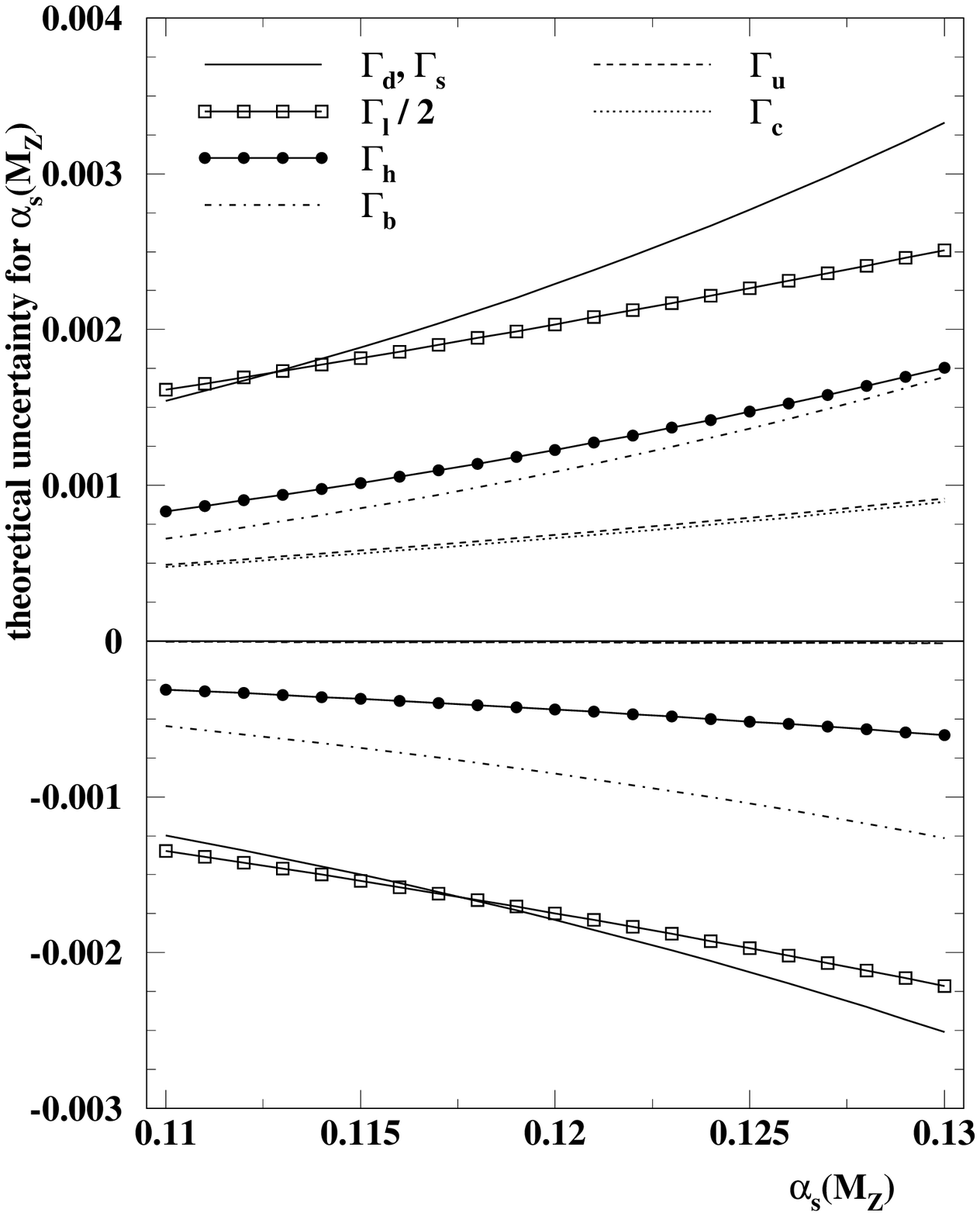, width=0.8\textwidth}
\caption{Positive and negative contributions to the
 perturbative systematic uncertainty of $\alpha_s(\mz)$ determined from 
various partial widths of the Z boson. The uncertainty using $\Gamma_l$ 
obtained at NLO is 
divided for representation by a factor of two.\label{fig:Zaswi}}}

As expected from the scale uncertainty of the width itself, there 
are large differences between the uncertainties of $\alpha_s$ determined 
using the widths of up-, down-type quarks and leptons. The size of the uncertainty 
for $\alpha_s$ is between one and two percent for quarks and about 4 percent 
for leptons, QCD corrections for the leptonic widths being calculated only at NLO. 
 Given the shape of the 
scale uncertainty for the width of the u- and c-quark, the resulting uncertainty 
of $\alpha_s$ is essentially one-sided. Also for the other widths  
a certain asymmetry in the uncertainty is observed, the positive (upward) 
uncertainty is generally larger than the negative (downward). This asymmetry may well 
be a technical artefact of the scale variation prescription, and conservatively 
the maximum of the positive and negative uncertainty is assigned 
as a symmetric uncertainty. For selected input values of $\alpha_s(\mz)$ the symmetrised uncertainties are 
given in Table~\ref{tab:aswidths}. 
\TABULAR[h]{|c|c|c|c|c|c|}{
\hline
$\alpha_s(\mz)$ &  $0.110$  & $0.115$   & $0.120$   & $0.125$   & $0.130$ \\ \hline
$\Gamma_u$      & $0.00049$ & $0.00058$ & $0.00068$ & $0.00079$ & $0.00092$ \\ 
$\Gamma_{d,s}$  & $0.00154$ & $0.00189$ & $0.00229$ & $0.00277$ & $0.00333$ \\ 
$\Gamma_c$      & $0.00047$ & $0.00056$ & $0.00066$ & $0.00077$ & $0.00089$ \\ 
$\Gamma_b$      & $0.00066$ & $0.00085$ & $0.00109$ & $0.00136$ & $0.00170$ \\ 
$\Gamma_h$      & $0.00083$ & $0.00102$ & $0.00123$ & $0.00147$ & $0.00176$ \\
$\Gamma_l$      & $0.00323$ & $0.00363$ & $0.00407$ & $0.00453$ & $0.00502$\\ \hline
%$\Gamma_Z$      & $0.00140$ & $0.00162$ & $0.00189$ & $0.00222$ & $0.00258$ \\ 
%$\sigma^0_h$    & $0.00070$ & $0.00086$ & $0.00104$ & $0.00126$ & $0.00151$ \\
%$\sigma^0_l$    & $0.00076$ & $0.00093$ & $0.00112$ & $0.00135$ & $0.00161$ \\
%$R_Z$           & $0.00079$ & $0.00095$ & $0.00114$ & $0.00139$ & $0.00165$ \\ \hline
}
{\label{tab:aswidths}
Systematic perturbative uncertainties for measurements of $\alpha_s$ from 
EW observables. The symmetric uncertainty is given by the maximum of upward 
and downward uncertainties obtained by a renormalisation scale variation.}
The uncertainty of $\alpha_s$ determined from realistic electroweak observables is 
shown in Fig.~\ref{fig:Zasobs}.  
\FIGURE[h!]
{\epsfig{file=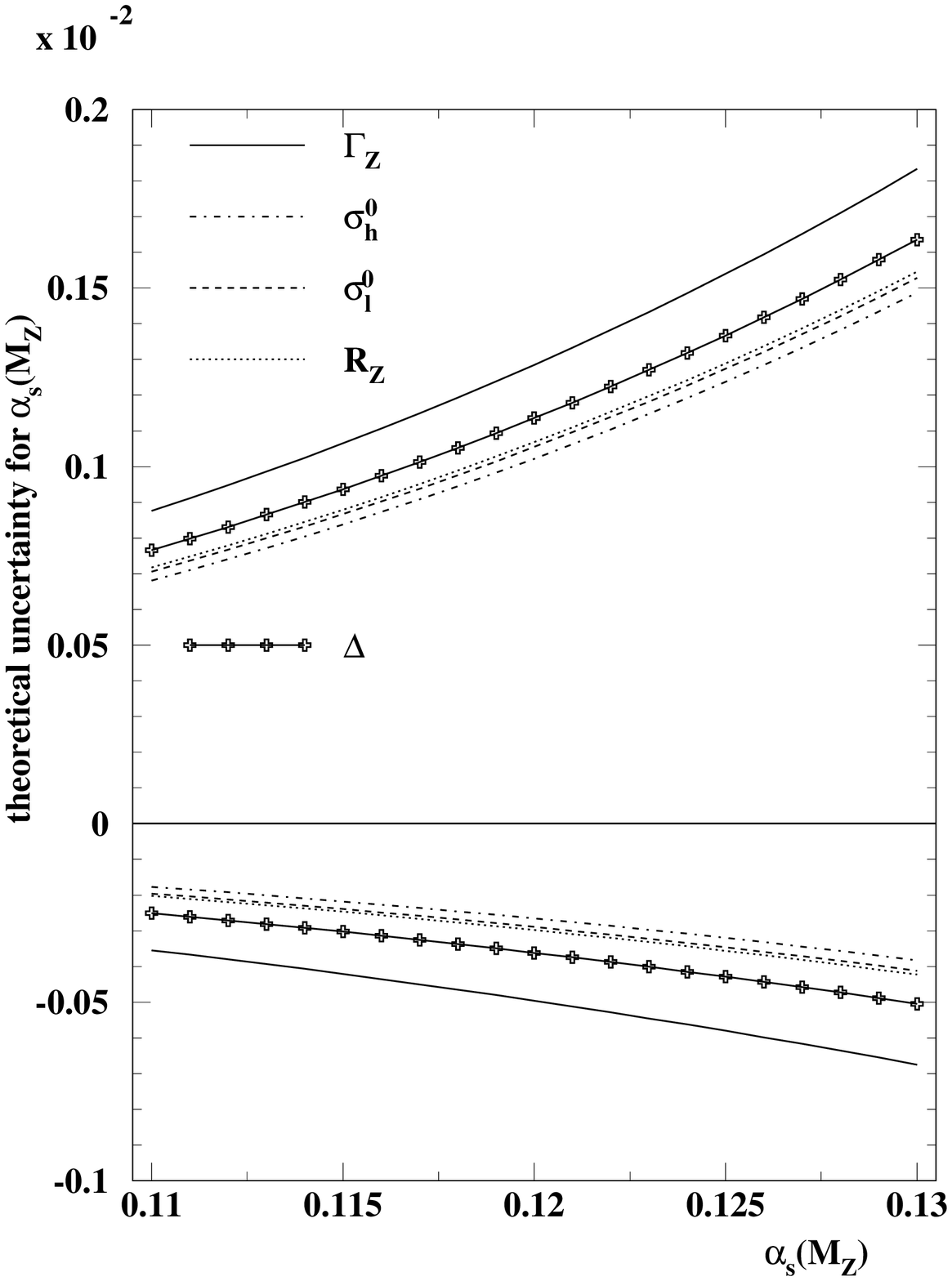, width=0.8\textwidth}
\caption{Systematic perturbative uncertainties for 
measurements of $\alpha_s$ using the EW observables $\Gamma_Z$, $R_Z$, $\sigma^0_l$ 
and $\sigma^0_h$ as function of the input value of $\alpha_s(\mz)$. The quantity 
$\Delta$ denotes a weighted average of the uncertainty for $\Gamma_Z$, $R_Z$ and 
$\sigma^0_h$, relevant for global analyses of electroweak data.\label{fig:Zasobs}}}
The perturbative uncertainty ranges between $1.3 \%$ for $\Gamma_Z$ and 
$1.0 \%$ for $\sigma^0_h$.  
The largest uncertainty is observed in the case of $\Gamma_Z$, which 
is directly proportional to the product of effective couplings and radiator functions. 
The uncertainties from the other variables are very similar, between $1.0 \%$ and $1.1 \%$.  
In $R_Z$ the widths appear linearly 
in nominator and denominator, in $\sigma^0_l$ and $\sigma^0_h$ quadratic combinations 
of leptonic, hadronic and total widths of the Z boson interplay. As a consequence 
the scale dependence of the effective couplings and/or the radiator functions 
cancel to some extent in the ratio. 

Several observables are included in the global EW fits \cite{EWG}, 
but only $R_Z$, $\Gamma_Z$ and $\sigma^0_h$ have a sizeable sensitivity 
to $\alpha_s$. Effectively, these three observables determine $\alpha_s$ and the 
perturbative uncertainty of $\alpha_s$ is bound to be an average of their individual 
contributions. The exact weights of each variable to the determination of $\alpha_s$ 
and subsequently to its perturbative uncertainties can not be determined in the 
framework of the present work. In order to derive nonetheless an estimate 
of the perturbative uncertainty from a global fit, a new variable 
\begin{equation}
\Delta=\omega_1 \Gamma_Z + \omega_2 \sigma^0_h + \omega_3 R_Z \; ,
\end{equation} 
is introduced in order to approximate the sensitivity of the three 
real observables to $\alpha_s$ in a global fit by a single observable. 
The perturbative uncertainty for $\alpha_s$ obtained from the $\Delta$ variable 
is an indicative for the true uncertainty from a combined fit.  
The weights $\omega_i$ are determined from the sensitivity 
\begin{equation}
\omega_i = \left|\frac{\partial O_i}{\partial \alpha_s}\right| \frac{\sigma(\alpha_s)}{\sigma(O_i^{exp})} \; ,
\end{equation}
where $O_i$ are the three observables, $\sigma(\alpha_s)$ is the uncertainty on $\alpha_s$ from 
the fit, taken to be $0.003$ for all observables, and $\sigma(O_i^{exp})$ is the 
measured experimental uncertainty for the observables themselves. This procedure 
yields $\omega_1=0.36$, $\omega_2=0.23$ and $\omega_2=0.41$. The resulting uncertainty 
from $\Delta$ is also shown in Fig.~\ref{fig:Zasobs}. 
The upper and lower bounds of the combined uncertainty can be estimated 
assuming that the 
true weights of the observables in a global fit are all positive. In this case 
the combined uncertainty can not be larger or smaller than any of 
the individual contributions from the single observables.   

In Table \ref{tab:asobs} the symmetric perturbative uncertainties are summarised for 
the realistic observables and for the $\Delta$ variable in a narrow range 
around $0.119$. The possible variations for the combined uncertainty 
are also given as $\Delta^+$ for the upper and $\Delta^-$ for the lower bound. 
As a stability test, the weights in the calculation of $\Delta$ have been set constant 
for all observables. The resulting combined uncertainty does not change by 
more than $0.00001$. 

\TABULAR[h]{|l|c|c|c|c|c||c|c|}{
\hline
$\alpha_s(\mz)$ &  $0.118$  & $0.119$   & $0.120$   & $0.121$   & $0.122$  & $a$ & $b$\\ \hline
$\Gamma_Z$      & $0.00119$ & $0.00124$ & $0.00128$ & $0.00133$ & $0.00138$ &$-1.22\cdot10^{-4}$ & $6.80$ \\ 
$\sigma^0_h$    & $0.00095$ & $0.00098$ & $0.00102$ & $0.00106$ & $0.00110$ &$-1.70\cdot10^{-4}$ & $5.77$ \\
$\sigma^0_l$    & $0.00098$ & $0.00101$ & $0.00105$ & $0.00110$ & $0.00114$ &$-1.60\cdot10^{-4}$ & $5.88$ \\
$R_Z$           & $0.00099$ & $0.00103$ & $0.00107$ & $0.00111$ & $0.00115$ &$-1.55\cdot10^{-4}$ & $5.92$  \\ \hline
$\Delta$        & $0.00105$ & $0.00109$ & $0.00114$ & $0.00118$ & $0.00122$ &$-1.50\cdot10^{-4}$ & $6.22$ \\
$\Delta^+$      & $0.00119$ & $0.00124$ & $0.00128$ & $0.00133$ & $0.00138$ & &  \\
$\Delta^-$      & $0.00095$ & $0.00098$ & $0.00102$ & $0.00106$ & $0.00115$ & &  \\ \hline
}
{\label{tab:asobs} Systematic perturbative uncertainties for measurements of $\alpha_s$ from 
EW observables. The symmetric uncertainty is given by the maximum of upward 
and downward uncertainties obtained by a renormalisation scale variation. The variable $\Delta$ 
represents the combined uncertainty for a global fit including $\Gamma_Z$, $\sigma^0_h$ and $\sigma^0_l$, 
with upper bound $\Delta^+$ and lower bound $\Delta^-$.}
For a value of $\alpha_s(\mz)=0.12$ the perturbative uncertainty ranges from $\pm 0.00102$ for $\sigma^0_h$ 
to $\pm 0.00128$ for $\Gamma_Z$, with a weighted average of $\pm 0.00115$ (unweighted $\pm 0.00114$). 

The dependence of the symmetric uncertainty on 
the input value of $\alpha_s(\mz)$ can smoothly be 
parameterised to form $a+b\cdot \alpha_s(\mz)^4$. The parameterisation, valid for  
$0.11 \leq \alpha_s(\mz) \leq 0.13$, allows for an interpolation of the uncertainty 
between calculated points. The parameters $a$ and $b$ are given for each observable 
in Table \ref{tab:asobs}. The result of the parameterisation is compared in 
Fig.\ref{fig:Zaspar} to the exact calculation, which is reproduced to good accuracy.   
\FIGURE[h!]
{\epsfig{file=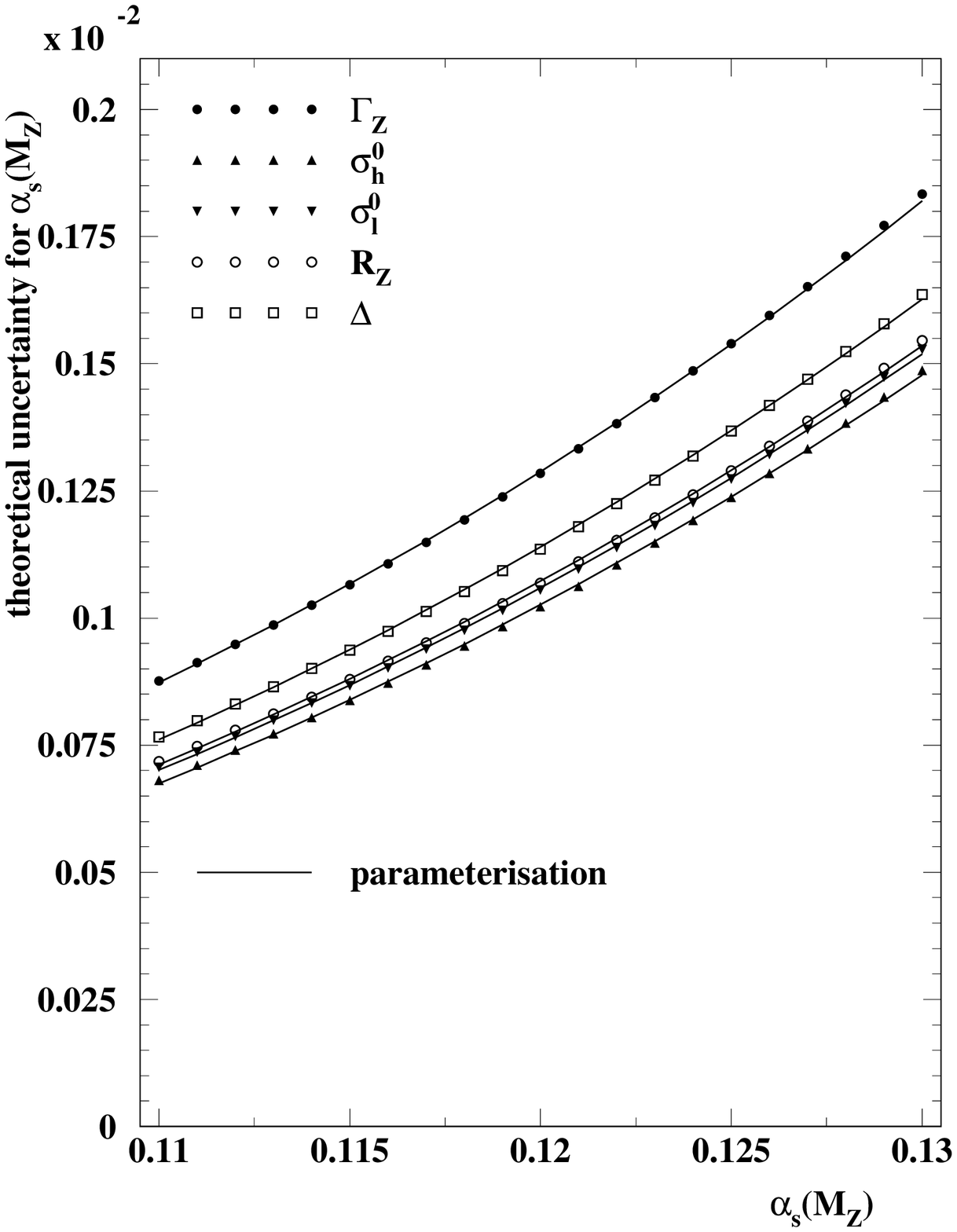 , width=0.8\textwidth}
\caption{Symmetrised systematic perturbative uncertainties for 
measurements of $\alpha_s$ using the EW observables. The quantity 
$\Delta$ denotes an weighted average of the uncertainty for $\Gamma_Z$, $R_Z$ and 
$\sigma^0_h$, relevant for global analyses of electroweak data. The symbols show the 
exact calculation and the lines represent the result of the parameterisation.\label{fig:Zaspar}}}

\subsection{Experimental tests}
\label{sec:etest}
In order to substantiate the uncertainty estimate, a fit is performed to 
a set of test data taken from the preliminary LEP combination \cite{EWG}. 
In the fit to the test data, the value of $\alpha_s$ is determined 
for $\xmu=1$ (nominal case) and $\xmu^+ , \xmu^-$ for the uncertainty. 
Only realistic observables are considered and the other SM parameters are fixed 
to the values given in Section~\ref{sec:ewobs}. The central values, 
experimental systematic uncertainties and correlations between the observables 
used are given in Table~\ref{tab:ewtest} (taken from \cite{EWG}).    
\TABULAR{|l|cc|ccc|}{
\hline
observable & central & experimental & \multicolumn{3}{|c|}{correlations} \\ 
           & value   & uncertainty  &  $\Gamma_Z$ & $\sigma^0_h$ & $R_Z$  \\ \hline
$\Gamma_Z [\rm{GeV}]$  & $2.4952$ & $0.0043$ & $1.0$ & $-0.297$ & $0.004$ \\ 
$\sigma^0_h [\rm{nb}]$ & $41.540$ & $0.037$  &        & $1.0$    & $0.183$ \\
$R_Z$                  & $20.767$ & $0.024$  &        &          & $1.0$  \\
$\sigma^0_l [\rm{nb}]$ & $2.0003$ & $0.0027$ &   &   &   \\ \hline
}
{\label{tab:ewtest}
Test data for the EW observables. The numerical values are preliminary results taken 
from \cite{EWG}, since $\sigma^0_l$ is not included in the global fit its correlation data is not available.}
A fit to the test data yields results for $\alpha_s$ summarised in Table~\ref{tab:ewfit}. 
The central values for fits using single observables are close to the world average \cite{Bethke1} 
apart from $\sigma^0_h$ which yields 
a very low value of $\alpha_s$ with a large experimental uncertainty of $6\%$. As expected, 
the smallest error of $3\%$ is obtained for $\sigma^0_l$ having the best sensitivity. 
The perturbative uncertainties obtained in this test are in good agreement with 
the data-independent method summarised in Table~\ref{tab:asobs}. The case of $\sigma^0_h$ 
yielding a small theoretical uncertainty originating from its small central result illustrates 
the bias appearing when the uncertainty is determined from an actual measurement. This bias 
can be reduced by taking a combined world average for $\alpha_s$ in the theoretical 
estimation of the perturbative uncertainty. 

The experimental precision of fits to several observables is increased to $3\%$ for the canonical 
ensemble of $\Gamma_Z $, $\sigma^0_h$ and $R_Z$ which is also used by the global EW fits \cite{EWG}. 
The perturbative uncertainty is again consistent with the result obtained previously 
for the $\Delta$ variable. 
The observables $\sigma^0_h$ with less sensitivity may be exchanged by $\sigma^0_l$ 
in the data sample to be fit. The correlation of experimental uncertainties between    
$\sigma^0_l$ and the other variables is unknown, as a crude approximation the same correlation 
as for $\sigma^0_h$ is assumed. Under this reserve the experimental uncertainty is 
further reduced to  $2\%$ with essentially unchanged perturbative uncertainties. 
As the experimental uncertainty dominates the precision for $\alpha_s$, the replacement 
of $\sigma^0_h$ by $\sigma^0_l$ might be considered for the global EW fits.     

\TABULAR{|l|cc|ccc|}{
\hline
included & central & experimental & $\xmu^+$ & $\xmu^-$ & perturbative  \\ 
observable    & value   & uncertainty  &         &          & uncertainty   \\ \hline
$\Gamma_Z $   & $0.1174$ & $0.0041$ & $0.1170$ & $0.1186$ & $-0.00043 +0.00122$ \\ 
$\sigma^0_h$ & $0.1076$ & $0.0065$ & $0.1074$ & $0.1082$ & $-0.00015 +0.00063$ \\
$R_Z$         & $0.1231$ & $0.0037$ & $0.1228$ & $0.1244$ & $-0.00033 +0.00125$  \\
$\sigma^0_l$  & $0.1187$ & $0.0030$ & $0.1184$ & $0.1197$ & $-0.00027 +0.00104$   \\ \hline
$\Gamma_Z $, $\sigma^0_h$, $R_Z$    & $0.1191$ & $0.0027$ & $0.1188$ & $0.1203$ & $-0.00043 +0.00113$ \\ 
$\Gamma_Z $, $\sigma^0_l$, $R_Z$ $)^*$    & $0.1202$ & $0.0021$ & $0.1199$ & $0.1213$ & $-0.00031 +0.00114$ \\ \hline
}
{\label{tab:ewfit}
Fit results for $\alpha_s(\mz)$ from fits to various test data sets. The central result 
is obtained for $\xmu=1$, the perturbative uncertainty from a variation for the renormalisation 
scale from $\xmu^+$ to $\xmu^-$. $)^*$ In the combined fit to $\Gamma_Z $, $\sigma^0_l$ and $R_Z$ 
the correlation for $\sigma^0_l$ is assumed to be the same as for $\sigma^0_h$.}
The LEP electroweak working group has adopted another strategy 
to incorporate the QCD uncertainties in the global fits. 
From the perturbative uncertainties for the observables $\Gamma_Z $, $\sigma^0_h$ and $R_Z$ 
themselves, given in Table~\ref{tab:obserr}, a covariance matrix was constructed and 
added to the other covariance matrices related to statistical and further systematic 
uncertainties. The total covariance matrix is then included in the global 
fit. While the central value of $\alpha_s$ does not change significantly when 
the QCD covariance matrix is added, its total uncertainty does increase and 
by quadratic subtraction of the non-QCD uncertainties a perturbative uncertainty 
of $\pm 0.0010$ is obtained \cite{Gruenewald}.  
This result confirms the estimate of perturbative uncertainty based on theoretical considerations 
for the $\Delta$ variable given in Table~\ref{tab:asobs}.

\section{Conclusions}
\label{sec:conc}
%============================================================================
A new method has been presented for the perturbative uncertainties at NNLO 
of measurements of $\alpha_s$ obtained from global analyses of precision electroweak 
data. The systematic uncertainties are obtained by a variation of the renormalisation scale  
in the calculations of final state QCD and mixed QCD$\otimes$EW corrections for 
the electroweak observables included in the global analyses used to determine 
$\alpha_s$ and other Standard Model parameters. 
The NNLO massless corrections, quark mass corrections and mixed 
QCD$\otimes$EW corrections are observed to contribute to the 
theoretical predictions for the observables at the 
level of one per mil. 
Individual contributions 
to the renormalisation scale dependence have been studied in detail and the 
resulting uncertainty has been calculated for the widths of the Z boson into 
different quark flavours and leptons. For electroweak observables used 
to constrain the Standard Model and determine $\alpha_s(\mz)$ the 
corresponding perturbative uncertainty for a value of $\alpha_s(\mz)=0.12$ is estimated 
to be between $\pm 0.0010$ and $\pm 0.0013$, in average $\pm 0.0011$. 
The size of the perturbative uncertainty has been cross-checked with a toy fit 
to experimental test data. 
The determination of $\alpha_s$ at NNLO from electroweak data is one 
of most precise measurements of $\alpha_s(\mz)$. Its precision is 
dominated by experimental effects yielding a relative uncertainty of about $3\%$, 
while perturbative uncertainties contribute only $1\%$.         
    
%============================================================================

\section*{Acknowledgements}
\begin{samepage}
The author would like to S.~Bethke who has initially raised the issue of 
the perturbative uncertainties in this context. The author is indebted to    
G.P.~Salam and R.~Harlander for providing the essential ingredient on the renormalisation 
scale variation in presence of running quark masses. This article would not have been 
completed without the expertise provided by M.~Gr\"unewald on the topic of precision 
electroweak measurements and constraints on the Standard Model.  
\end{samepage}

\end{document}

%% file: defs.tex
\def\text{\textstyle}

\newcommand{\bq}{\begin{equation}}
\newcommand{\eq}{\end{equation}}
\newcommand{\ba}{\begin{eqnarray}}
\newcommand{\ea}{\end{eqnarray}}
\newcommand{\nl}{\nonumber\\}
\newcommand{\nll}{\nonumber\\}

\newcommand{\nf }{\mbox{$n_f  $}}
\newcommand{\nfS}{\mbox{$n^2_f$}}

\newcommand{\mqS}{\mbox{$m^2_q$}}
\newcommand{\mqQ}{\mbox{$m^4_q$}}
\newcommand{\mqX}{\mbox{$m^6_q$}}

\newcommand{\mqpQ}{\mbox{$m'^4_q$}}
\newcommand{\ms }{\mbox{$m_s  $}}

\newcommand{\mc }{\mbox{$m_c  $}}
\newcommand{\mcS}{\mbox{$m^2_c$}}
\newcommand{\mcQ}{\mbox{$m^4_c$}}

\newcommand{\mbS}{\mbox{$m^2_b$}}
\newcommand{\mbQ}{\mbox{$m^4_b$}}

\newcommand{\Mc }{\mbox{$M_c  $}}

\newcommand{\MqS}{\mbox{$M^2_q$}}

\newcommand{\McS}{\mbox{$M^2_c$}}
\newcommand{\MbS}{\mbox{$M^2_b$}}

\newcommand{\LMSBn }{\mbox{$\Lambda^{(\nf)}_{\overline{\mathrm{MS}}}$}}
\newcommand{\LMSBnS}{\mbox{$\left(\Lambda^{(\nf)}_{\overline{\mathrm{MS}}}
\right)^2$}}
\newcommand{\LMSBnml }{\mbox{$\Lambda^{(\nf-1)}_{\overline{\mathrm{MS}}}$}}

\newcommand{\als }{\alpha_{_s}}

\newcommand{\ztwo}{\zeta(2)}
\newcommand{\ztri}{\zeta(3)}
\newcommand{\zfor}{\zeta(4)}
\newcommand{\zfiv}{\zeta(5)}
\def\srt{\sqrt{2}}

\def\s0h{\sigma^0_h}
\def\nl{\nonumber \\}
\newcommand{\bqa}{\begin{eqnarray}}
\newcommand{\eqa}{\end{eqnarray}}

\newcommand{\sss}[1]{\scriptscriptstyle{#1}}

\newcommand{\gf}{G_{\mu}}

\newcommand{\zb}{Z}

\newcommand{\ff}{f}

\newcommand{\fl}{l}
\newcommand{\fq}{q}

\newcommand{\mzs}{M^2_{\sss{Z}}}
\newcommand{\mzc}{M^3_{\sss{Z}}}

\newcommand{\mts}{m^2_{t}}

\newcommand{\mhl }{M_{\sss{H}}}

\newcommand{\mzl }{M_{\sss{Z}}}

\newcommand{\lrbr}{\left[}
\newcommand{\rrbr}{\right]}

\newcommand{\qf  }{Q_f  }

\newcommand{\siw }{s_{\sss{W}}}           

\newcommand{\siws}{s^2_{\sss{W}}}

\newcommand{\vc }[1]{v_{#1}}
\newcommand{\ac }[1]{a_{#1}}

\newcommand{\Rvaz}[1]{g^{#1}_{\sss{\zb}}}

\newcommand{\tcif}{I^{(3)}_f}

\newcommand{\sman}{s}

\newcommand{\smans}{s^2}

\newcommand{\Reb}{{\rm{Re}}}

\newcommand{\rZf}{\rho^{\ff}_{\sss{Z}}}
\newcommand{\rZl}{\rho^{\fl}_{\sss{Z}}}
\newcommand{\rZq}{\rho^{\fq}_{\sss{Z}}}
\newcommand{\gZf}{g^{\ff}_{\sss{Z}}}
\newcommand{\gZl}{g^{\fl}_{\sss{Z}}}
\newcommand{\gZq}{g^{\fq}_{\sss{Z}}}
\newcommand{\kZf}{\kappa^{\ff}_{\sss{Z}}}

\newcommand{\kZdf}[1]{\kappa^{#1}_{\sss{Z}}}

\newcommand{\gqq}{\Gamma_{\fq}}

\newcommand{\Imsi}[1]{I^2_{#1}}

\newcommand{\epem}   {\ensuremath{\mathrm{e^+e^-}}}

\newcommand{\mz}     {\ensuremath{M_{\rm Z}}}
\newcommand{\mw}     {\ensuremath{M_{\rm W}}}

\newcommand{\xmu}     {\ensuremath{x_\mu}}

\def\asb{{\overline \alpha_s}}

%% file: jhep2.bbl
\begin{thebibliography}{10}

%% actual references relative ZFITTER

\bibitem{EWG} The LEP Collaborations ALEPH, DELPHI, L3, OPAL, the LEP Electroweak 
working group, the SLD Electroweak and Heavy Flavour Working group, 
``\,{\it A Combination of preliminary Electroweak Measurements and constraints on the 
Standard Model}\,'', hep-ex/0412015

\bibitem{Hasko} R.~Jones, M.~Ford, G.P.~Salam, H.~Stenzel and D.~Wicke
``\,{\it Theoretical uncertainties on $\alpha_s$ from event-shape
variables in $e^+e^-$ annihilations}\,'', JHEP {\bf 12} (2003) 007.

\bibitem{Soper} D.E.~Soper and  L.R.~Surguladze, 
``\,{\it On the QCD perturbative expansion for $\epem \rightarrow$ hadrons}\,'', Phys. Rev.  {\bf D54} (1996) 4566.

\bibitem{Bethke1} S.~Bethke, Nucl. Phys. Proc. Suppl.  {\bf 135} (2004) 345. 
%``\,{\it On the QCD perturbative expansion for $\epem \rigtharrow$ hadrons}\,'', 

\bibitem{Bethke2} S.~Bethke, J. Phys.  {\bf G 26} (2000) R27. 
%``\,{\it On the QCD perturbative expansion for $\epem \rigtharrow$ hadrons}\,'', 



\bibitem{zfitter} D.~Bardin et al., ``\,{\it ZFITTER v.6.21 - A Semi-Analytical Program for Fermion Pair Production in $\epem$ Annihilation}\,'', Comput. Phys.\ Commun.\ {\bf 133} (2001) 229.

\bibitem{czak1}
A.~Czarnecki and J.~H. K{\"u}hn, Phys. Rev. Lett. {\bf 77} (1996)
  3955. %--3958.

\bibitem{czak2}
R.~Harlander, T.~Seidensticker, and M.~Steinhauser, Phys. Lett. {\bf
  B426} (1998) 125.%--132.

\bibitem{deltalep} M. Steinhauser, Phys. Lett. {\bf B429} (1998) 158. 

\bibitem{deltahad} H. Burkhardt and B. Pietrzyk, Phys. Lett. {\bf B513} (2001) 46.  

\bibitem{Kniehl} B.~A.~Kniehl, Nucl. Phys. {\bf B347} (1990) 86. 

\bibitem{zphys}
M.~Consoli, W.~Hollik, and F.~Jegerlehner, ``{\em Electroweak radiative corrections
  for {$Z$} physics}'', in {\em Proc. of Workshop on Z Physics at LEP, Geneva,
  Switzerland, Feb 20-21 and May 8-9, 1989, {\rm report CERN 89--08 (1989)}}

\bibitem{afmt}
L.~Avdeev, J.~Fleischer, S.~Mikhailov, and O.~Tarasov, Phys. Lett. {\bf B336} 
(1994) 560, E: ibid., {\bf B 349} (1995) 597; \\
K.~G. Chetyrkin, J.~H. {K\"uhn}, and M.~Steinhauser, Phys. Lett. {\bf B351} (1995) 331. %--338.
\bibitem{ChKS}
K.~G. Chetyrkin, J.~H. {K\"uhn}, and M.~Steinhauser, Phys. Rev. Lett. {\bf 75}, (1995) 
3394. %--3397.

\bibitem{zbb}
A.~Akhundov, D.~Bardin, and T.~Riemann, ``{\em Electroweak one loop corrections to
  the decay of the neutral vector boson}'', Nucl. Phys. {\bf B276} (1986)
  1.

\bibitem{ftjr}
J.~Fleischer, O.~V. Tarasov, F.~Jegerlehner, and P.~Raczka, Phys. Lett.
  {\bf B293} (1992) 437. %--444.

\bibitem{taub2}
J.~Fleischer, O.~V. Tarasov, and F.~Jegerlehner, Phys. Lett. {\bf B319}
  (1993) 249. %--256.

\bibitem{Chetyrkin:1979bj}
K.~G. Chetyrkin, A.~L. Kataev, and F.~V. Tkachev, Phys. Lett. {\bf B85}
  (1979) 277.

\bibitem{Dine:1979qh}
M.~Dine and J.~Sapirstein, Phys. Rev. Lett. {\bf 43} (1979) 668.

\bibitem{Celmaster:1980xr}
W.~Celmaster and R.~J. Gonsalves, Phys. Rev. Lett. {\bf 44} (1980) 560.

\bibitem{Gorishnii:1991hw}
S.~G. Gorishny, A.~L. Kataev, and S.~A. Larin, Phys. Lett. {\bf B273}
  (1991) 141. %--144.

\bibitem{Chetyrkin:1994js3}
K.~Chetyrkin, J.~K{\"u}hn, and A.~Kwiatkowski, ``{{QCD}} corrections to the
  $e^+ e^-$ cross-section and the {Z} boson decay rate'', in {\em Reports of
  the Working Group on Precision Calculations for the {$Z$} Resonance, {\rm
  report CERN 95--03 (1995)}} (D.~Bardin, W.~Hollik, and G.~Passarino, eds.),
  pp.~175--263.

\bibitem{Gruenewald} M.~Gr\"unewald, private communication.



\end{thebibliography}
